\documentclass[twocolumn,prb,showpacs]{revtex4}
\usepackage{psfrag}
\usepackage{graphicx}
\usepackage{amsmath}
\usepackage{amssymb}
\usepackage{eufrak}
\usepackage[ansinew]{inputenc}

\begin{document}

\title{Neutron to proton mass difference, parton distribution functions and baryon resonances from dynamics on the Lie group u(3)}

\author{Ole L. Trinhammer}
\affiliation{Department of Physics, Technical University of Denmark, \\
Fysikvej, Building 307, DK-2800 Kongens Lyngby, Denmark.\\
ole.trinhammer@fysik.dtu.dk } 

\begin{abstract}
We present a hamiltonian structure on the Lie group u(3) to describe the baryon spectrum. The ground state is identified with the proton. From this single fit we calculate approximately the relative neutron to proton mass shift to within half a percentage of the experimental value. From the same fit we calculate the nucleon and delta resonance spectrum with correct grouping and no missing resonances. For specific spin eigenfunctions we calculate the delta to nucleon mass ratio to within one percent. Finally we derive parton distribution functions that compare well with those for the proton valence quarks. The distributions are generated by projecting the proton state to space via the exterior derivative on u(3). We predict scarce neutral flavour singlets which should be visible in neutron diffraction dissociation experiments or in invariant mass spectra of protons and negative pions in B-decays and in photoproduction on neutrons. The presence of such singlet states distinguishes experimentally the present model from the standard model as does the prediction of the neutron to proton mass splitting. Conceptually the Hamiltonian may describe an effective phenomenology or more radically describe interior dynamics implying quarks and gluons as projections from u(3) which we then call allospace.
\end{abstract}

Short title: Baryons from u(3)

\maketitle

\section*{1 Introduction}
The quark flavour model has a missing resonance pro\-blem since it predicts many more baryon 
$\pi N$-resonances than observed [1, 2, 3]. The quark colour model (QCD) has a confinement problem to construct hadrons analy\-tically from quarks and gluons. We are aware of the successes of QCD in pertubative domains [4, 5] and in lattice gauge theory [6, 7]. Nevertheless we want to stress the two first mentioned issues. The latter, the confinement problem, motivated a radical approach which solves the former as a by-product. We construct the dynamics in a compact space. The spectroscopy can live there and manifest itself in real space as different mass resonances. Then we will have confinement per construction and can hope to see quarks and gluons by projection from the compact configuration space to the laboratory space. We will present here a study of this idea. We shall call it the allospatial hypothesis from the greek word {\it{allos}} meaning another or different. In general terms the idea is that a hamiltonian description is the more natural framework for spectroscopy and the lagrangian description more suitable for scattering phenomena, see e.g.\ [8].

The defining equation is a group space Hamiltonian on u(3) where the toroidal degrees of freedom project out in colour quark fields and gluons come out by an adjoint projection. It turns out that there is also room for flavour in the model and we can reproduce an Okubo mass formula involving hypercharge and isospin. We note from the beginning that u(3) is not to be thought of as a symmetry group but as a configuration space.

The allospatial hypothesis is evolved in sect. 2. Here a specific Hamiltonian is stated, the underlying quantizations defined and a projection of the group space wave function into real space quark and gluon fields given. The projection is shown in appendix A to lead to fields transforming properly as respectively fundamental and adjoint representations of su(3). Appendix A concludes: "From the projection \ldots \ to laboratory space we recognize the toroidal generators as momentum operators. Thus when experimental production of resonances is of concern we see from space: {\it{The impact momentum generates the (abelian) maximal torus of the u(3) allospace. The momentum operators act as introtangling generators.}} When decay, asymptotic freedom, fragmentation and confinement is of concern we see from allospace: {\it{The quark and gluon fields are projections of the vector fields induced by the momentum form}} $d \Phi$." With this interpretation of the origin of quarks we see that the quarks are confined per construction since the Lie group of allospace is compact and thus cannot be projected in its global totallity. "You cannot peel an orange without breaking the skin", as differential topologists say, see fig. 1.

The theory is unfolded in sect. 3. The Laplacian is parametrized in a polar decomposition analogous to the treatment of the hydrogen atom in polar coordinates. We shall see that the Laplacian contains a term naturally interpreted as a centrifugal potential. It includes the off-toroidal generators of the group and thereby carries the complexity needed to include spin and flavour spectral characteristics alongside the basic colour dynamics. As an example we calculate matrix elements of our Hamiltonian with symmetrized $D$-functions of specific angular momentum and find a very promising ratio when compared with $m_{\Delta(1232)}/m_{N(939)}$. 

Energy eigenstates are discussed in sect. 4 and both approximate and exact solutions given for the neutral flavour N- and $\Delta$-sectors together with tentative spin-parity assignments. The number and grouping of resonances agrees with all the certain (four star) resonances listed by the particle data group without a missing resonance problem.

In sect. 5 we give a controversial interpretation rela\-ting period doublings in the wavefunction to the creation of charge in the neutron decay. The mass shift related to this interpretation is quite promising with a relative mass difference of 0.13847 \% between the neutron and proton mass predicted from the approximate solutions to be compared with the experimental value of 0.13784 \%. In sect. 6 we discuss experimental predictions. Apart from predicting all the observed, certain resonances our model predicts scarce neutral flavour singlets. In sect. 7 we list some open questions together with selected parton distribution functions and in sect. 8 we give concluding remarks. 

The present work contains five appendicies. In appendix A we give the projection to space. In appendix B we derive the spectrum of the off-toroidal generators of our group space Laplacian. In appendix C we describe the Rayleigh-Ritz method used to solve the exact case for neutral charge states and  calculate analytically the matrix elements of the Rayleigh-Ritz method when this is used on trigonometric base functions. In appendix D we derive parton distribution functions. In appendix E we give a survey of those concepts in differential geometry most crucial for the calculations in appendicies A and D.

\begin{figure}
\begin{center}
\includegraphics[width=0.35\textwidth]{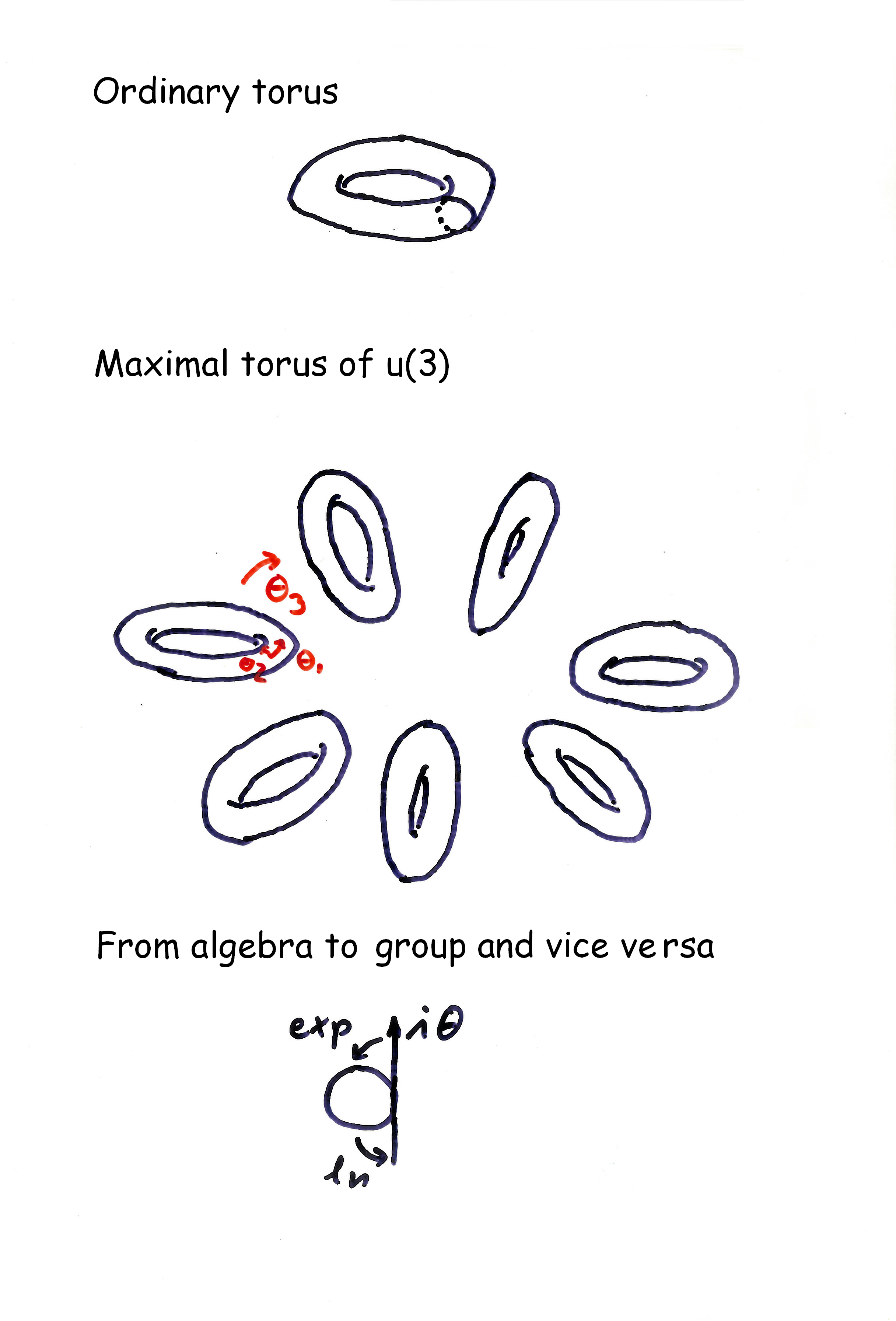}
\caption{The algebra approximates the group in the neighbourhood of origo. The allospace in the present work is the Lie group u(3) in which the dynamics is supposed to describe the baryon spectrum as a set of stationary states of a certain Hamiltonian. Projecting from the group to the algebra gene\-rates quark and gluon fields according to 	the representation space chosen.}
\label{fig1}
\end{center}
\end{figure}

\section*{2 The allospatial hypothesis}

The 'communication' between the interior dynamics and the spacetime dynamics runs through the exterior derivative of the state on the chosen configuration space. So the characteristics of the fields we wish to generate by the exterior derivative implies which configuration space should be hypothesized. We wish to generate projection fields transforming under the SU(3) algebra with the fields possibly being electrically charged. This points to a configuration space containing both su(3) and u(1). Thus we choose the Lie group u(3) as configuration space and assume the following Hamiltonian
\begin{equation}
   \frac{\hbar c}{a} \left[ - \frac{1}{2} \Delta + \frac{1}{2} d^2(e,u)\right] \Psi(u) = E \Psi (u).
\end{equation}
It is the hypothesis of the present work, that the eigenstates of (1) describe the baryon spectrum with $u \in u(3)$   being the configuration variable of a sole baryonic entity and $a$ is a scale. Below we shall find exact solutions of (1) for alleged N-states and we shall discuss approximate solutions for both alleged N- and $\Delta$-states. It should be mentioned that, when unfolded, the structure of (1) carries degrees of freedom for both colour, spin, hypercharge and isospin.	

The Laplacian $\Delta$  in (1) is parametrized in the next section. The configuration variable $u=e^{i\chi}$, and the trace of its squared argument $Tr\chi^2=d^2(e,u)$  is used in the representation independent potential. So the potential is half the squared geodetic distance [9] from the 'point' $u$ to the 'origo' $e$
\begin{equation}    
d^2(e,u) = \theta^2_1 + \theta^2_2 + \theta^2_3, \, \, \, - \pi \le \theta_j \le \pi , 
\end{equation}		
where $e^{i\theta_j}$  are the eigenvalues of $u$. The geodetic distance may be seen as the euclidean measure folded into the group manifold [10]. The geodetic distance is invariant under translation in group space as it should be since the choice of origo is arbitrary
\begin{equation}    
   d(v,uv) =d(e,v^{-1}uv)=d(e,u).
\end{equation}

\begin{figure} [h]
\begin{center}
\includegraphics[width=0.35\textwidth]{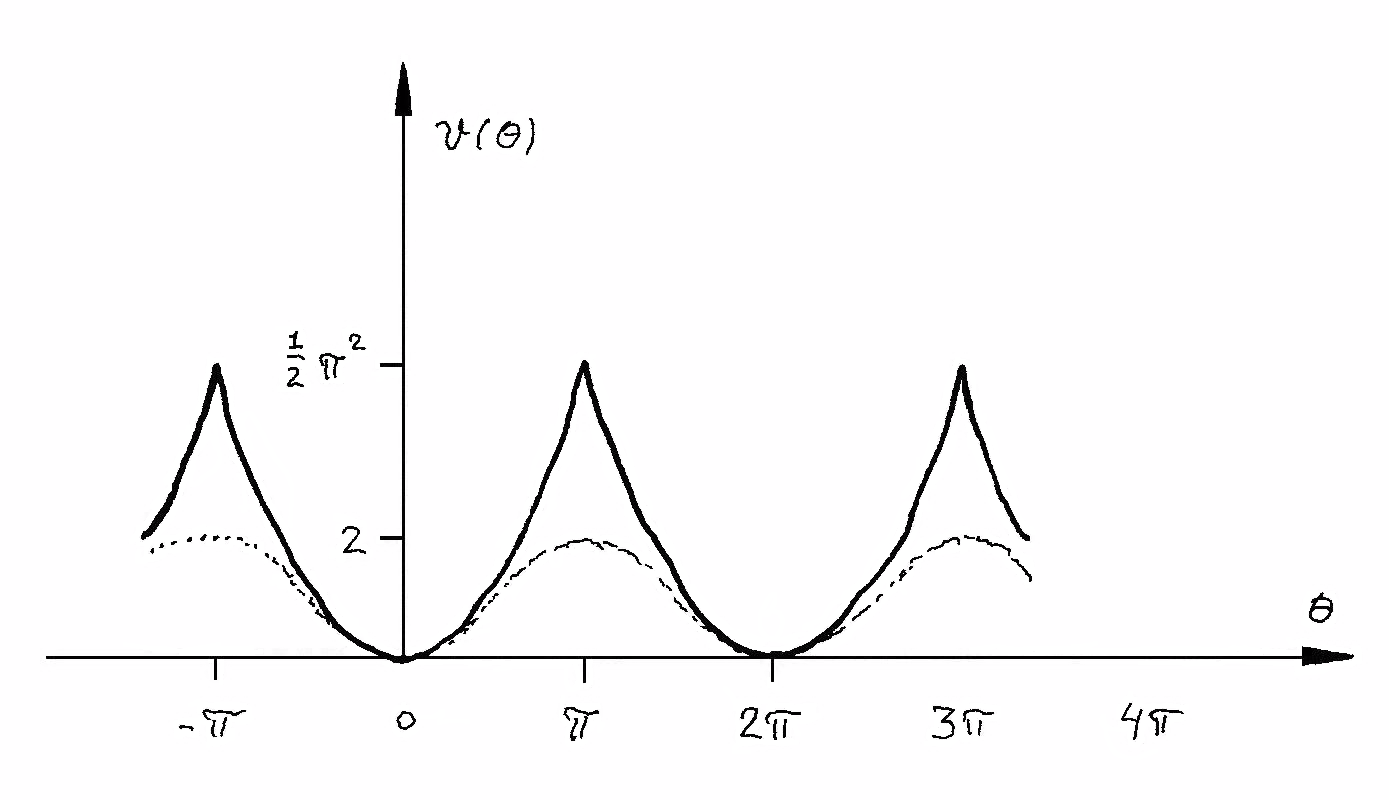}
\caption{Periodic parametric potential (17) originating from the squared geodetic distance (2) from $u$ to the origo $e$ in the Lie group (Milnor [9]). The dashed curve corresponds to the Wilson analogue [11] of the Manton potential [12] and is not considered in the present work.}
\label{fig2}
\end{center}
\end{figure}
	
Planck's constant $h=2 \pi \hbar$  enters the description via the canonical quantization behind eq. (1). The speed of light $c$ and the scale $a$ enter the description from a parametrization of the eigenangles $\theta$  via either a time parameter $t$ or a space parameter $x=a \theta = ct$. This gives the scale
$\hbar \omega=\hbar \dot{\theta} =\hbar c/a  $   for the dimensionful eigenva\-lues $E$. The ground state eigenvalue E $= E/(\hbar c/a) $  of the dimensionless edition (10) of (1) is of the order of 4.5 predicting a promising scale $a\approx$ 1 fm if the dimensionful ground state eigenvalue is identified with the nucleon $E$ = 939 MeV. Reversing the argument $a$ is to be determined experimentally by fixing the ground state energy of (1) to that of the nucleon. It turns out that $\Lambda \equiv \hbar c / a $= 210 MeV in fine agreement with the average QCD scale, e.g.\ $\Lambda^{(5)}_{\overline{M}\, \overline{S}}$  in [13].

We denote the configuration variable $u$ by a lower case letter and accordingly the Lie group space to make it clear that $u$ is a configuration variable and not a priori a symmetry transformation. We parametrize the confi\-guration variable by  $u=e^{i \alpha_kT_k}$ where $T_k$ for $k = 4,\ldots ,9$ are the off-toroidal Gell-Mann generators (14), whereas $T_j$ for $j$ = 1,2,3 are toroidal generators and  $\alpha_j = \theta_j$. 

We have toroidal coordinate fields  $\partial_j$ implied by the (abelian) toroidal generators
\begin{equation}    
     \partial_j =  \frac{\partial}{\partial \theta} ue^{i\theta T_j} \vert _{\theta=0} = u i T_j.
\end{equation}
The basic action-angle quantization condition ($\hbar=1$) with $T_j = -i \partial/ \partial \theta_j = - i \partial_j\vert_e$  reads
\begin{equation}    
[ \partial_j, \theta_i ]= \delta_{ij} \Leftrightarrow d \theta_i (\partial_j) =  \delta_{ij},
\end{equation}
for $i, j$ = 1,2,3, where $d  \theta_i$  are the torus forms.

The scale $a$ enters the model when the three toroidal degrees of freedom are related to the three spatial degrees of freedom by the following projection
\begin{equation}    
   x_j = a \theta_j .
\end{equation}
Thus we define parametric momentum operators  $p_j$  proportional to the toroidal generators
\begin{equation}    
   p_j = - i \hbar \frac{1}{a} \frac{\partial}{\partial \theta_j} = \frac{\hbar}{a} T_j,
\end{equation}				
whereby the canonical action-angle quantization of (5) translates into position-momentum quantization
\begin{equation}    
   \left[ a \theta_i,p_j \right] = - i \hbar \delta_{ij}, \, \, \, \, \, 
\left( \left[ a \theta_i , a \theta_j \right] =   \left[ p_i , p_j \right] =0 \right).
\end{equation}
It is via (6) and (8) that the physical dimensions enter the model (1).

The state $\Psi$ projects out on laboratory space via $d\Phi$ where  $\Phi = J \Psi$ and $J$ is the Jacobian (13) of our parametrization. For instance the restriction of the momentum form $d\Phi$  to the torus
\begin{equation}    
     d \Phi = \psi_jd\theta_j
\end{equation}						
generates a member of the fundamental representation of SU(3). The coefficients in (9) are the components of the alleged colour quark spacetime field whose transformation properties are shown in appendix A.

\section*{3 The theory unfolded}

To study the stationary states of (1) let us rewrite the Schr\"{o}dinger equation in a dimensionless form with  
E = $E/ \Lambda$
\begin{equation}    
    \left[ - \frac{1}{2} \Delta +  \frac{1}{2}d^2 (e,u) \right] \Psi(u) = {\rm{E}} \Psi(u).
\end{equation}
The Laplacian from a polar decomposition on u(3) is [14]
\begin{equation}    
     \Delta = \sum^3_{j=1} \frac{1}{J} \frac{\partial^2}{\partial \theta^2_j} J
+ 2 - \! \! \! \sum^3_{\substack{ i <  j \\ k\neq i,j}} \frac{K^2_k + M^2_k}{8 \sin^2 \frac{1}{2}(\theta_i -\theta_j)}. 
\end{equation}
We recognize a term for toroidal kinetic energy and two terms which may be interpreted in the Hamiltonian as additional potentials: a constant global curvature [15] potential $\frac{1}{2} \cdot (-2)=-1$  and a centrifugal potential
\begin{equation}    
     C= \frac{1}{2}  \sum^3_{\substack{ i <  j \\ k\neq i,j}} \frac{K^2_k + M^2_k}{8 \sin^2 \frac{1}{2}(\theta_i -\theta_j)}. 
\end{equation}
Further the van de Monde determinant, the 'Jacobian' of our parametrization is
\begin{equation}    
      J = \prod^3_{i <  j } 2 \sin\left(\frac{1}{2} (\theta_i- \theta_j)\right),
\end{equation}
and with  $\hbar = 1$ the off-diagonal Gell-Mann generators are [16, 17, 18, 19]
\begin{align}    
        K_1 = \lambda_7 , \quad & K_2 = \lambda _5, \quad   K_3 = \lambda_2 , \\
        M_1 = \lambda_6 , \quad & M_2 = \lambda_4 ,  \quad  M_3 = \lambda_1. \nonumber
\end{align}
$K_k$ commute as body fixed angular momentum operators and $M_k$ 'connect' the algebra by commuting into the subspace of $K_k$
\begin{equation}    
     \left[M_k,M_l \right] = \left[K_k,K_l \right]=-iK_m	\quad {\rm{(cyclic \, in}} \, k,l,m).
\end{equation}
The presence of the components of ${\bf{K}}=(K_1, K_2,K_3)$  in the Laplacian operator opens for the description of spin. Interpreting $\bf{K}$ as the allospatial spin operator is encouraged by the body fixed signature of the commutation relations. The relation between space and allospace is like the relation in nuclear physics between fixed coordinate systems and intrinsic body fixed coordinate systems for the description of rotational degrees of freedom. Once $\bf{K}^2$ and e.g.\ $K_3$ have been chosen to be independent and mutually commuting operators, both commuting with the Hamiltonian H of (1), the presence of  ${\bf{M}}=(M_1, M_2,M_3)$  in the Laplacian makes it possible to describe flavour via the spectrum of ${\bf{M}}^2$, see below.

Substituting (11) into (10) we get the fully parametrized Schr\"{o}dinger equation
\begin{align}    
{\large{[}} - \frac{1}{2}\left( \sum^3_{j=1} \frac{1}{J} \frac{\partial^2}{\partial \theta^2_j} J
+ 2 - \! \! \! \sum^3_{\substack{ i <  j \\ k\neq i,j}} \frac{K^2_k + M^2_k}{8 \sin^2 \frac{1}{2}(\theta_i -\theta_j)} \! \right)  \nonumber \\ 
 + v(\theta_1)+ v(\theta_2)+ v(\theta_3) \!  
 {\huge{]}}\Psi (u)= \rm{E} \Psi (u),
\end{align}
where the potential $\frac{1}{2}d^2$ spells out as a sum of periodic 'chopped' harmonic oscillator potentials. The potential (17) is shown in fig. 2,
\begin{align}    
   v(\theta)     = \frac{1}{2}(\theta - n  \pi)^2, \quad \theta \in [(2n-1)\pi,(2n+1)\pi], \nonumber \\
   {\rm{where}}\quad n\in Z. \hspace{4cm}
\end{align}

We may assume the wavefunction to be an eigenstate of {{\bf{K}}$^2$ and  {{\bf{M}}$^2$ and thus write it as a product of a toroidal part $\tau($\mbox{\boldmath$\theta$}$)$ like the radial wavefunction for the hydrogen atom and an off-toroidal part $\Upsilon_{KM}$  like the spherical harmonics $Y_{lm}$. With $\mbox{\boldmath$\theta$}  =(\theta_1, \theta_2, \theta_ 3)$ we thus write
\begin{equation}    
  \Psi(u)=\tau (\theta_1, \theta_2, \theta_ 3) \cdot \Upsilon_{KM}(\alpha_4,\alpha_5 ,\alpha_6 ,\alpha_7 ,\alpha_8 ,\alpha_9).
\end{equation}
It should be noted that in determining the spectrum for {{\bf{M}}$^2$ we exploit the possibility of choosing specific eigenstates of hypercharge and isospin 3-component as shown in appendix B and thus instead of $\Upsilon_{KM}$  we might had written $\Upsilon_{KYI_3}$. Multiplying (16) by the Jacobian $J$ we introduce a new function
\begin{equation}    
  \Phi(u)=R(\mbox{\boldmath$\theta$}) \cdot \Upsilon_{KM} \quad {\rm{where}}\quad R(\mbox{\boldmath$\theta$}) = J(\mbox{\boldmath$\theta$}) \cdot \tau (\mbox{\boldmath$\theta$}).
\end{equation}
Then $\Phi$ satisfies
\begin{gather}    
   [- \Delta_e - 2 +  \sum^3_{\substack{ i <  j \\ k\neq i,j}} \frac{K^2_k + M^2_k}{8 \sin^2 \frac{1}{2}              (\theta_i -\theta_j)} + 2 \sum^3_{j=1} v(\theta_j)]\, \Phi(u)  \nonumber \\
 = 2 {\rm{E}} \Phi(u),\hspace{3cm}\end{gather}
where the 'euclidean' Laplacian is
\begin{equation}    
 \Delta_e = \sum^3_{j=1} \frac{\partial^2}{\partial \theta_j^2}.
\end{equation}
Now we integrate out the off diagonal degrees of freedom $(\alpha_4,\alpha_5 \ldots,\alpha_9)$  to get an equation for the torodial part $R_{KM}$  for specific eigenvalues $K(K+1)$  and $M^2$  of respectively {\bf{K}}$^2$ and {\bf{M}}$^2$
\begin{equation}    
  [-\Delta_e +V] R_{KM}(\theta_1, \theta_2,\theta_3) = 2 {\rm{E}} R_{KM} (\theta_1, \theta_2,\theta_3).
\end{equation}
Here the curvature and centrifugal terms of the group space Laplacian have been collected with the geodetic distance potential into a total potential $V$
\begin{gather}    
  V = -2 + \frac{1}{3}(K(K+1)) + M^2) \sum^3_{ i <  j} \frac{1}{8 \sin^2 \frac{1}{2}(\theta_i -\theta_j)}
\nonumber \\\hspace{1cm} +  2 (v(\theta_1) + v(\theta_2)+ v(\theta_3)). \hspace{2.5cm}
\end{gather}
In the centrifugal term during the integration we have exploited the existence of the Haar measure over  $(\alpha_4,\alpha_5 \ldots,\alpha_9)$ together with the factorization (18). Further we have used that the off-toroidal part of the wavefunction is an eigenstate of {\bf{K}}$^2$ and {\bf{M}}$^2$ together with the fact that the centrifugal term (12) is symmetric under interchange of the torus angles $\theta_j$. 

The centrifugal term leads to a mass formula of the well-known Okubo type [20]. In appendix B we derive the spectrum of  {\bf{K}}$^2$ + {\bf{M}}$^2$   and show the following relation among quantum numbers
\begin{equation}    
  K(K+1) + M^2 = \frac{4}{3}\left(n+\frac{3}{2}\right)^2 - 3 - \frac{1}{3} y^2 - 4 i^2_3.
\end{equation}
Here $y$ is the hypercharge, $i_3$ is the three-component of isospin and $n$ is an integer which we may call hyperdimension. It is natural in the present framework to classify the eigenstates according to the three independent values of $n$, $y$ and $i_3$. However we can make a transformation of this classification into the familiar one by rewriting the expression (24) and choose the sum of hyperdimension and hypercharge to be a constant. For $n + y =2$, which yields the lowest possible  $K(K+1) + M^2$, we get
\begin{equation}    
  K(K+1) + M^2 \!=\! \frac{40}{3}+(k^2_3 + m^2_3) - \frac{28}{3}y + 4\left[ \frac{1}{4}y^2\! - i(i+1)\right].
\end{equation}
Since $(K^2_3 + M^2_3)$  commutes with both $Y$ and $I^2$  we get for a given value of  $(k^2_3 + m^2_3)$
\begin{equation}    
   K(K+1) + M^2 = a' + b' y +c'\left[\frac{1}{4}y^2 - i(i+1)\right].
\end{equation}

Equation (26) is the famous Okubo mass formula that reproduces the Gell-Mann, Okubo, Ne'eman mass relations within the baryon N-octet and  $\Delta$-decuplet [17, 20, 21, 22] independently of the values of $a'$, $b'$ and $c'$. Of course this is only so if one chooses the same toroidal wavefunction for all members of a given multiplet. In practice the SU(3) symmetry breaking in (26) will be influenced by the  $\mbox{\boldmath$\theta$}$-dependence in (23) because different values of {\bf{K}}$^2 +$ {\bf{M}}$^2$ lead to different values of the centrifugal potential and thereby influence which span of toroidal energy eigenstates will project out on a specific angular momentum eigenstate in the laboratory. 

\subsection*{3.2 Total angular momentum}

All members of an SU(3) multiplet have the same total angular momentum and therefore it would seem natural to let this be carried by the structure of the toroidal wavefunction. As a possible base one may try $D$-functions. The $D$-functions are eigenfunctions of angular momentum [23]
\begin{equation}    
  {\text{\bf{J}}}^2 D_{k,m}^j = j (j+1)D_{k,m}^j.
\end{equation}

Now the $D$-functions are functions of the Euler angles describing rotations between a fixed coordinate system and body fixed coordinates [24]. They are simultaneously eigenfunctions of the three-component of angular momentum in coordinate space and in body fixed space with eigenvalues $-m$  and  $-k$ respectively, where $m$ and $k$ may vary in integer steps from $-j$ to $j$ [25, 26]. Despite the notation no a priori relation is intended between $k$, $m$ and {\bf{K}}, {\bf{M}}. In a specific example below however, we shall test the identification $j=K$. In the present connection we may interpret the fixed coordinate system as the laboratory space and the body fixed space as allospace. Since a priori there is no prevailing direction in laboratory space the states should have no specific three-component of angular momentum. Thus states should be constructed by summing over $m$. And since the labeling of the eigenangles in allospace is arbitrary, we should expand on states that are symmetric under interchange of the eigenangles. Therefore a possible expression for the nucleon ground state wave function with $j = k =1/2$ could be
\begin{align}    
   T^{1/2}_{1/2}  (\theta_1, \theta_2, \theta_3) = (t^{1/2}_{1/2}  (\theta_1, \theta_2, \theta_3) + t^{1/2}_{1/2}  (\theta_2, \theta_3, \theta_1) \nonumber \\
+ t^{1/2}_{1/2}  (\theta_3, \theta_1, \theta_2))\! /N_{1/2}, \hspace{1.7cm}
\end{align}
where $N_{1/2}$ is a normalization factor and $t^{1/2}_{1/2}$  is a sum over the two $D$-functions for the two possible values 1/2 and -1/2  of $m$ corresponding to total angular momentum  $j=1/2$
\begin{gather}    
  t^{1/2}_{1/2} = D^{1/2}_{1/2,1/2} + D^{1/2}_{1/2,-1/2} \propto e^{- i \theta_1/2}\cos(\frac{\theta_2}{2})e^{- i \theta_3/2} \nonumber \\\hspace{-2.6cm}- e^{- i \theta_1/2}\sin(\frac{\theta_2}{2})e^{ i \theta_3/2}.\end{gather}

For total angular momentum $j=3/2$   there will be four terms in (29) corresponding to the allowed values of  $m = \frac{3}{2}, \frac{1}{2}, -\frac{1}{2}, -\frac{3}{2}$. It is encouraging for this line of investigation to note that
\begin{equation}    
 \frac{<JT^{3/2}_{3/2} \mid H_R \mid JT^{3/2}_{3/2}>}{<JT^{1/2}_{1/2} \mid H_R \mid JT^{1/2}_{1/2}>} = 1.32 \approx \frac{m_{\Delta(1232)}}{m_{N(939)}}= 1.31.
\end{equation}
Here H$_R$  is the allospatial Hamiltonian in the exact form in (22) and the result in (30) is for the lowest possible value of {\bf{K}}$^2 +$ {\bf{M}}$^2$  like in table 1. The value of $<JT^{1/2}_{1/2} \mid H_R \mid JT^{1/2}_{1/2}>$  in (30) is 5.93 and fitting this to the nucleon rest energy 939.6 MeV corresponds to a scale $a \approx 1.2$ fm. It is also reassuring to note that the numerical eigenvalue 5.93 is rather close to the value 0.5111/0.0855 = 5.98  of the nucleon ground state in a favoured choice of parameters for lattice calculations shown in table 5 of [7].

\begin{figure}
\begin{center}
\includegraphics[width=0.45\textwidth]{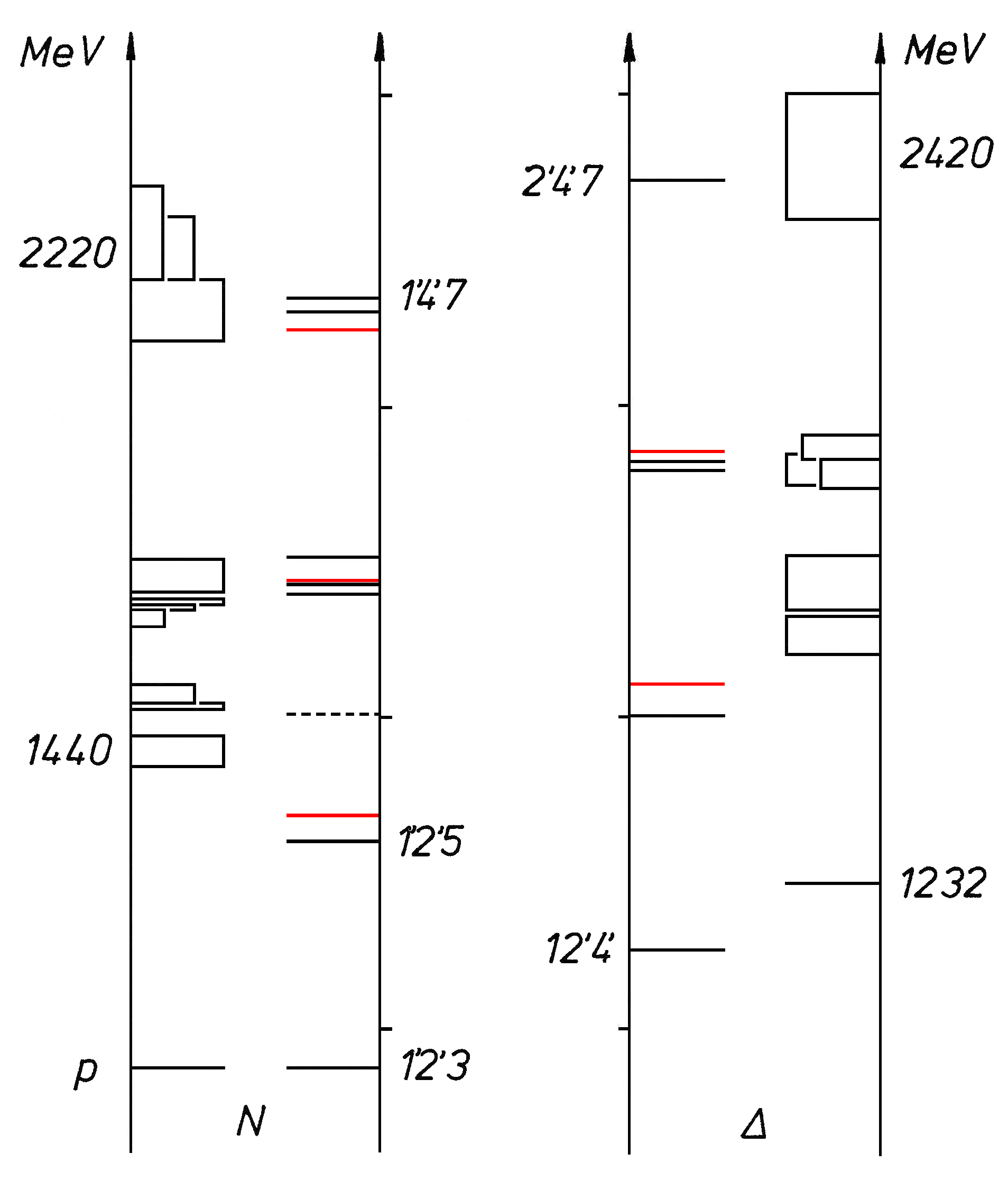}
\caption{All observed certain neutral flavour baryons (boxes) compared with approximate allospatial predictions (lines) from eq. (31) with label combinations based on table 2. The dashed line is a predicted neutral charge state without charged partners. The red lines mark states with augmented contribution in level 3. The boxes indicate the experimental range of peak values [27], not the widths which are much larger. The systematic errors in the approximation are of the order of 50 MeV. Digits at selected allospatial states are the toroidal labels $l,m,n$. Note the fine agreement in the grouping and the number of resonances in both sectors with just one fitting at $p=1',2',3$.}
\label{fig3}
\end{center}
\end{figure}

\subsection*{3.3 Energy eigenstates}}

Equation (22) can be solved by a Rayleigh-Ritz method, for details see appendix C. The eigenvalues of the full equation (22) are listed in table 1 for the lowest possible value of {\bf{K}}$^2 +$ {\bf{M}}$^2$. Here we have expanded $R(\mbox{\boldmath$\theta$})$  on Slater determinants of trigonometric functions of increasing order (44). In table 4 consistency is shown by comparison with exact results for (22) by expansions on Slater determinants of parametric eigenfunctions (33). In table 1 are shown also possible identifications with baryon resonances. We see that all the N-resonances which the Particle Data Group considers certain [27] are present except N(1520). Thus the general tendency is promising but the predicted energy values are in need of improvement. Such an improvement is seen if instead of comparing the energy of the electrically neutral partners of each N-resonance one looks at the energy of the charged partner. The energies of the charged partners are calculated below for an approximate case and the comparison is shown in fig. 3. In sect. 6 we shall return to the question of N(1520).

\vspace{4mm}

\section*{4 Approximate solutions}

Here we omit the constant curvature potential and the centrifugal potential. They are of the same order of magnitude and of opposite sign, see sect. 5. We may then expect to keep the general structure of the spectrum. We shall introduce a new degree of freedom to be related to the electric charge.

Having integrated out the off-toroidal degrees of freedom in (19), we are lead to an approximate edition of (22), namely the following seperable equation

\begin{gather}    
 \left[- \frac{1}{2} \sum^3_{j=1} \frac{\partial^2}{\partial \theta^2 _j} + v(\theta_1)+  v(\theta_2) + v(\theta_3)\right]  R(\theta_1, \theta_2,\theta_3)  \nonumber \\
 = {\rm{E}} R (\theta_1, \theta_2,\theta_3). \hspace{3cm}
\end{gather}
With both the Laplacian and the potential now being just sums, the equation separates into three Schr\"{o}dinger equations for parametric eigenfunctions with periodic boundary conditions
\begin{equation} 
 \left[ - \frac{1}{2} \frac{\partial^2}{\partial \theta^2 _j} + \frac{1}{2} \theta^2_j \right]\! \phi_i(\theta_j) = e_i \phi_i (\theta_j), \hspace{2.7mm} - \pi \le \theta_j \le \pi.
\end{equation}

\begin{figure} [h]
\begin{center}
\includegraphics[width=0.35\textwidth]{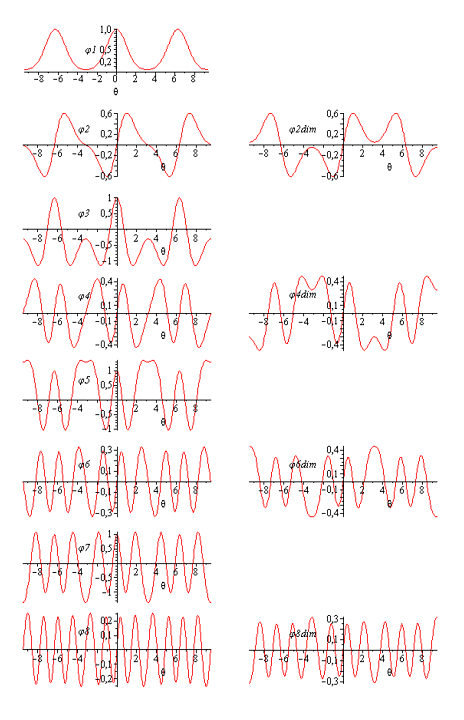}
\caption{Periodic parametric eigenfunctions (32) from which the toroidal part (34) of the wavefunction of approximate states are constructed. Eigenstates of diminished eigenvalues are shown to the right. The exact functions in (22) can be found from an expansion on such states or on a base set (44) mimicking these parametric eigenfunctions by sines and cosines of increasing order.}
\label{fig4}
\end{center}
\end{figure}

Figure 4 shows to the left the first eight parame\-tric eigenfunctions. The eigenstate $\Psi$  in (18) is symme\-tric in the eigenangles as the labelling of angles in the parametrization is arbitrary. $J$ is antisymmetric. In total  $R (\theta_1, \theta_2,\theta_3)$ is antisymmetric and may be constructed [28] as a Slater determinant of three freely chosen ortho\-gonal ($l < m < n$) parametric eigenfunctions, 
\begin{equation} 
   R_{lmn} (\mbox{\boldmath$\theta$}) = \begin{vmatrix}
    \phi_l( \theta_1) & \phi_l( \theta_2)& \phi_l( \theta_3) \\
   \phi_m( \theta_1) & \phi_m( \theta_2)& \phi_m( \theta_3)\\
      \phi_n( \theta_1)& \phi_n( \theta_2)& \phi_n( \theta_3) \\ \end{vmatrix}.
\end{equation}
Substituting the $KM$-indices each state thus gets three toroidal labels $l,m,n$, the labels of the parametric eigenfunctions. The eigenvalues E of (31) are sums of three different parametric eigenvalues E = $\sum_{i=l,m,n} e_i$. For instance we will have E = $ e_1 + e_2 + e_3$   for

\begin{equation} 
   \tau_{123}(\mbox{\boldmath$\theta$})  = \frac{1}{J}\begin{vmatrix}
    \phi_1( \theta_1) & \phi_1( \theta_2)& \phi_1( \theta_3) \\
   \phi_2( \theta_1) & \phi_2( \theta_2)& \phi_2( \theta_3)\\
      \phi_3( \theta_1)& \phi_3( \theta_2)& \phi_3( \theta_3) \\ \end{vmatrix}.
\end{equation}
We may interpret the antisymmetry under interchange of the columns in the Slater determinant $R$ as a colour antisymmetry under interchange of the eigenangles of $u$. To keep a uniform probability measure in group space when seen from parameter space, it is $\Phi$  that we would like to project to space and not  $\Psi$. Because the external derivative is linear $d\Phi$  inherits the antisymmetry of $\Phi$ and carries it through to the projection space, e.g.
\begin{equation} 
   d\Phi \mid_{(\theta_2, \theta_1 , \theta_3)} = d(-\Phi \mid_{(\theta_1, \theta_2 , \theta_3)}) = - d\Phi \mid_{(\theta_1, \theta_2 , \theta_3)}.
\end{equation}

\subsection*{4.2 Parity}

We might expect a 'winding number' effect in $\tau_{KM}$  due to the increasing number of oscillations in the base functions. Such a winding number effect may be expressed in specific partial waves to which the allospatial state relates in space. This would explain the general trend of increasing total angular momentum with increasing energy. The idea is supported by table 5 where we see the allospatial states together with spin and parity of the N-resonance candidates. We can give an ad hoc assignment of parity according to each particular combination of toroidal labels $l,m,n$, namely
\begin{equation} 
   P = (-1)^{m-l} \cdot(-1)^{n-m},
\end{equation}
which in general reduces to $ P = (-1)^{n-l} $. The assignment (36) gives a rather consistent correlation to the observed resonances with only a few exceptions namely for the singlet states like 135 (see below) which seems to mix with ordinary doublet N-states to yield a state with reversed parity. There also seems to be a correlation between increments in toroidal label number and increments in angular momentum, a tentative trend being $\Delta J = \Delta m + \Delta n$ in series like $125, 127, 147 \rightarrow \frac{1}{2} +, \frac{5}{2} +, \frac{9}{2} +$ and $134, 136, 156 \rightarrow \frac{1}{2} -, \frac{5}{2} -, \frac{9}{2} -$. This confirms the winding number effect. Using the parity assignment (36) and the trend $\Delta J = \Delta m$  we predict from the series $235, 237, (257) \rightarrow \frac{3}{2} -, \frac{7}{2} -,( \frac{11}{2} -) $  a $J^P$ value of $\frac{11}{2} -$ for the allospatial state 257 with approximate energy 2542 MeV in accordance with the  three star rated N(2600) [27]. Although promising the spin-parity assignments in table 5 is just a tentative number game. We shall now give a more specific suggestion.

We return to the projection $a\theta_j = x_j$  and construct parity eigenstates
\begin{equation} 
   T^+ (\mbox{\boldmath$\theta$})\!=\! T(\mbox{\boldmath$\theta$}) + T(-\mbox{\boldmath$\theta$}) \hspace{1mm} {\rm{and}} \hspace{1mm}  T^- (\mbox{\boldmath$\theta$})\! =\! T(\mbox{\boldmath$\theta$}) - T(-\mbox{\boldmath$\theta$})
\end{equation}
based on states like (28). Identifying the operation of the parity operator P on {\bf{x}} with a similar operation on the allospace eigenangles $\mbox{\boldmath$\theta$}$ would ensure that the states $T^{\pm}$   have intrinsic parity  $\pm 1$
\begin{equation} 
   {\rm{P}}(T^{\pm}(\theta)) \equiv T^{\pm}(-\theta) = \pm T^{\pm}(\theta).
\end{equation}
Since [P,$\,$H] = 0 states are possible with simultaneous $P$ and $E$ eigenvalues. A state like $T^{1/2}_{1/2}$  in (28) is then split into two with specific parity. The presence of two different parity states for each angular momentum fits very well with the pattern of the observed certain N-resonances. Such states might be the spin-parity-definite states through which the spin-parity-indeterminate energy eigenstates of parametric function Slater determinants like (33) communicate with the partial wave amplitudes of the experimental analysis.

\vspace{4mm}

\begin{figure} [h]
\begin{center}
\includegraphics[width=0.5\textwidth]{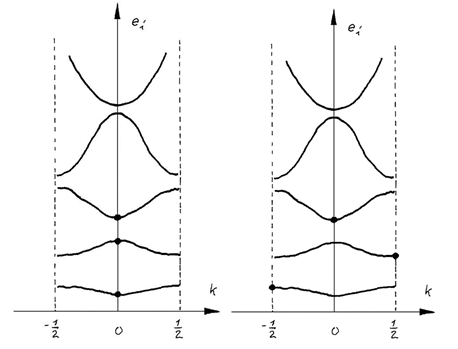}
\caption{Reduced zone scheme [29] for allospatial parametric eigenvalues. The black dots represent the values for the unstable neutron state (left) and the proton state (right). For clarity the variation of the eigenvalues with wave number   is grossly exaggerated for the lowest states.}
\label{fig5}
\end{center}
\end{figure}

\subsection*{4.3 Bloch states}

Now let us return to the question of solving (31) and (32). The basic trick in exploring the full structure of  $u(3)$ and solving the parametric eq. (32) is to introduce Bloch waves $\phi(\theta) = e^{i \kappa \theta}g(\theta)$  and get a set of eigenvalue bands, see fig. 5. Here $g$ is the  $2 \pi$ periodic part belonging to the Bloch wave number $\kappa$. Choosing  $\kappa = 0$ gives us $2 \pi$  periodicity of the parametric functions corresponding to the periodicity of the parametric potential. For the state with lowest eigenvalue in this case E = $e_1 + e_2 +e_3 = 4.474172455$, see table 6. If we choose $\kappa_{2'} = \pm \frac{1}{2}$ in level 2 it will lead to a lower lying parame\-tric eigenstate which we can couple to a 'compensating'  $\kappa_{1'} = \mp \frac{1}{2}$ in level 1. We thus couple two Bloch waves $\phi_{1'}(\theta) =e^{i \kappa_1 \theta}g_{1'}(\theta)$  and $\phi_{2'}(\theta) =e^{i \kappa_2 \theta}g_{2'}(\theta)$  from the two lowest lying eigenvalue bands in the reduced zone scheme to the right in fig. 5. This combination is possible because the energy gained by the period doubling in level 2 is larger than the energy lift needed to cause period doubling in level 1. Note that the parametric eigenstates of the first and second levels now have $4 \pi$  periodicity in the eigenangles. This is possible since we only require $\Psi^2$ and not nescessarily $\Psi$  to be single valued on $u(3)$ after a single "rotation" in the parameters $\theta_j$  (cf.\ the case of a spin $\frac{1}{2}$ particle like the free electron which needs a $4 \pi$  rotation for its state to return to its original value [30]). Other values of $\kappa$  than 0 and $\pm\frac{1}{2}$ are not in accordance with the single valuedness of $\Psi^2$. The lower eigenvalue after this coupled decay is E$^{''} = e_1'+ e_2' +e_3 = 4.467985519$ which we interpret as the proton ground state. With this interpretation we predict from our approximate calculation the relative difference between the neutron mass $m_{\rm{n}}$ and the proton mass $m_{\rm{p}}$
\begin{gather} 
  \frac{m_{\rm{n}}-m_{\rm{p}}}{m_{\rm{p}}} = \frac{E - E^{''}}{E^{''}} =
  \frac{4.474172455 - 4.467985519}{4.467985519}  \nonumber \\= 0.138473 \, \%. \hspace{2.6cm}\end{gather}
The result in (39) is in close agreement with the experimentally determined value [27]
\begin{gather} 
    \frac{m_{\rm{n}}-m_{\rm{p}}}{m_{\rm{p}}} = \frac{939.565346 \, {\rm{MeV}} - 938.272013 \, {\rm{MeV}} }{938.272013 \, {\rm{MeV}}} \nonumber \\= 0.137842  \, \%. \hspace{2.1cm}\end{gather}
It is encouraging that the approximate calculation gives a too high mass splitting. One may namely expect the eigenvalue bands to narrow slightly in an exact solution because of the spikes from the centrifugal potential which has been omitted in our above calculation and which will effectively leave the levels in a deeper potential well. On the other hand one may fear that the resulting level shifts might be too large in order to keep the fine agreement indicated by (39) and (40). However, the neutron state shifts only two per cent downwards to 4.38 for the exact solution in table 1. Actually the Jacobian (13) 'kills' the wave function exactly at the location of the singularities of the centrifugal potential, so we may hope that the shifting effect on the proton state will be just what we would need in order to diminish the still significant discrepancy between the prediction in (39) and the very accurate observations in (40) where the uncertainties on both masses are $\pm 0.000023$ MeV which means that the uncertainty lies on the last digit in (40).

\section*{5 Charge interpretation and systematic error}

We interpret the period doublings in the diminished ground state topology as 'creator' of the proton charge in the neutron decay. Thus $\kappa$ is interpreted as a charge quantum number degree of freedom which express the coupling between allospatial structure and the spacetime fields of the electroweak interaction. The interpretation is supported by the structure of the implied proton state
\begin{align}  
  \hspace{-6cm} \tau_{1^{'}2^{'}3} (\mbox{\boldmath$\theta$})\! = \hspace{5.8cm} \nonumber \\
 \frac{1}{J} \begin{vmatrix}
   e^{-i\frac{1}{2} \theta_1}g_{1'}(\theta_1) & e^{-i\frac{1}{2} \theta_2}g_{1'}(\theta_2) &
   e^{-i\frac{1}{2} \theta_3}g_{1'}(\theta_3)\\
   e^{i\frac{1}{2} \theta_1}g_{2'}(\theta_1) & e^{i\frac{1}{2} \theta_2}g_{2'}(\theta_2) &
   e^{i\frac{1}{2} \theta_3}g_{2'}(\theta_3)\\
   \phi_3(\theta_1) & \phi_3(\theta_2) & \phi_3(\theta_3) 
\end{vmatrix}. \hspace{5mm} \end{align}
We see that the period doublings induce u(1) factors curled up in the Slater determinant thereby "trapping" the charge topologically. Note that u(1) is the gauge group of the electromagnetic interaction. The diminished parametric function in the second row of the determinant in (41) is shown to the right in fig. 4 and the lifted one in the first row is shown in fig. 6.

\begin{figure} [h]
\begin{center}
\includegraphics[width=0.4\textwidth]{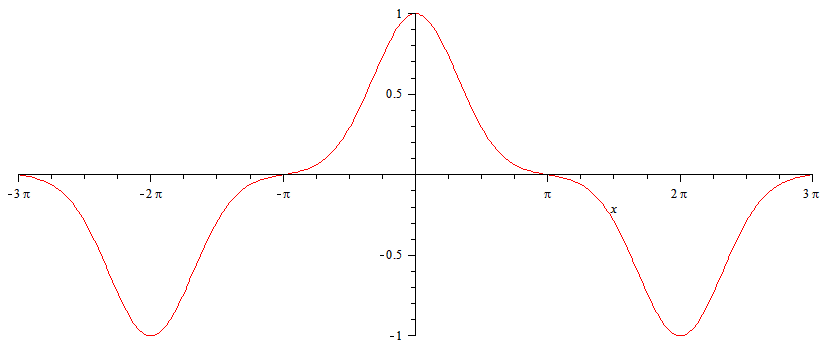}
\caption{The first lifted parametric eigenstate with  periodicity $4\pi$. The eigenvalue is found by collocation with 1500 points in Matlab and the corresponding eigenstate is generated by solving in Maple the one-dimensional Schr\"{o}dinger equation (32) for the particular eigenvalue. See [31] for programs.}
\label{fig6}
\end{center}
\end{figure}

The decay seems to couple inherently to spin degrees of freedom since the two Bloch wave phase factors render the measure scaled wave function
\begin{align} 
       R_{1'2'3} (\mbox{\boldmath$\theta$}) = \hspace{5.8cm} \nonumber \\
 \begin{vmatrix}
   e^{-i\frac{1}{2} \theta_1}g_{1'}(\theta_1) & e^{-i\frac{1}{2} \theta_2}g_{1'}(\theta_2) &
   e^{-i\frac{1}{2} \theta_3}g_{1'}(\theta_3)\\
   e^{i\frac{1}{2} \theta_1}g_{2'}(\theta_1) & e^{i\frac{1}{2} \theta_2}g_{2'}(\theta_2) &
   e^{i\frac{1}{2} \theta_3}g_{2'}(\theta_3)\\
   \phi_3(\theta_1) & \phi_3(\theta_2) & \phi_3(\theta_3) 
\end{vmatrix}, 
\end{align}with a structure resembling sums of $D$-functions [32]. For instance
\begin{align} 
   D^{\frac{1}{2}}_{\frac{1}{2},-\frac{1}{2}} (\theta_1, \theta_2, \theta_3) =  \hspace{4.0cm} \nonumber \\
  e^{-i\frac{1}{2} \theta_1} d^{\frac{1}{2}}_{\frac{1}{2},-\frac{1}{2}} (\theta_2) e^{i\frac{1}{2}           
\theta_3} = -e^{-i\frac{1}{2} \theta_1} \sin(\frac{1}{2} \theta_2) e^{i\frac{1}{2}\theta_3}\end{align}     
is related to a particular rotation of spinors for the case of spin $\frac{1}{2}$ [33, 34]. We may even imagine that the decay with its simultaneous period doublings in two parametric levels generates two component structures to carry away the excess energy. It is beyond the scope of the present work to show that these structures might be the electron and its antineutrino [35]. 

Instead of expansions on the parametric eigenfunctions (33) we could also use the Rayleigh-Ritz method for solving the full (22) and the approximate (31) eqs. by expansions on functions of the type 
\begin{align} 
   b_{pqr}(\theta_1, \theta_2, \theta_3) =
   \begin{vmatrix}
   \cos  p \theta_1 & \cos p \theta_2 & \cos  p \theta_3 \\
   \sin  q \theta_1 & \sin  q \theta_2 & \sin  q \theta_3 &\\
   \cos  r \theta_1 & \cos  r \theta_2 & \cos  r \theta_3
   \end{vmatrix}
   \nonumber \\
  = \epsilon_{ijk} \cos  p\theta_i \sin q \theta_j \cos r \theta_k, \hspace{2.0cm} \end{align}
where $p = 0,1,2, \ldots;  \, q = 1,2, \ldots; \, r = p+1, p+2, \ldots ;$ and similar with 0, 2 or 3 sines. The details are laid out in appendix C and it turns out that all matrix elements for this set of base functions can be calculated analytically. The functions (44) have a structure quite similar to the Slater determinant (33). The increasing number of oscillations in the parametric eigenfunctions of (33) is modeled by the increasing order of the cosines and sines.

The order of magnitude of the omitted terms in (31), i.e.\ the constant curvature potential and the centrifugal potential, may be evaluated from these base functions. For instance
\begin{equation} 
   <b_{011} \mid C \mid b_{011} \, > / < b_{011} \mid b_{011} > \, = \frac{2}{3},
\end{equation}
where $C$ is the centrifugal potential for $K^2 = \frac{1}{2}( \frac{1}{2}+1)$ and $M^2 = \frac{13}{4}$ like in table 1. The centrifugal potential more or less cancels the curvature potential -1  and the result  -1/3 of the two terms taken together is small compared to the geodetic potential which yields $\pi^2/2$, see (C12) and (C9).

\section*{6 Experimental predictions}

We determine the scale $\Lambda$  for the approximate solutions from the proton rest energy $ \Lambda = E / {\rm{E}} = 938.3$ MeV/4.468 = 210 MeV. Based on this scale the predictions are shown in fig. 3. The allospatial spectrum agrees with the number and grouping of all the certain resonances in the N- and $\Delta$-sector. It should be stressed that all the observed certain resonances are accounted for by allospatial states. By 'certain' we mean all the well established resonances with four stars in the particle data group listings [27]. Relative to the exact solutions in table 1 the omissions in the approximate solutions re\-present systematic errors of the order of 50 MeV.

The relative mass splitting between the neutron and the proton resulting from the period doublings in (39) is 0.13847 \% to be compared with the experimental value of 0.13784 \%. The slightly lighter (as opposed to heavier) mass of the charged proton is natural in the present hypothesis. States with two even labels like 1,$\,$2,$\,$4 have the possibility of two parametric eigenvalue diminishing period doublings. This will give states with a multiplicity of four. The state 1,$\,$2,$\,$4 is interpreted as the $\Delta$-resonance with two diminutions for $\Delta^{++}$. Again the charged partners are naturally the lighter ones. The state 2,$\,$4,$\,$6 is the first example where all three parametric states can undergo diminution induced period doublings. But the complex exponentials are now common factors in each column of the Slater determinant so they factorize out thereby releasing the topological trap. The determinant regains a parametric $2\pi$  periodicity and the state may eventually contribute to neutral resonances.

The grouping of the resonances agrees quite well with the observed certain states. Only one certain N-resonance is missing in the group of three resonances in the domain around 1500 MeV. However the approximate treatment in (31) predicts a neutral singlet 1,$\,$3,$\,$5 at 1510 MeV exactly in that area. The exact treatment in (22) places the singlet at 1526 MeV, see table 3. This state is thought to mix with the other two N-resonances to give the total of three N-resonances in the group. The next singlet state 1,$\,$3,$\,$7 is predicted in the 'desert' area between 1700 MeV and 2100 MeV. In the approximate case of (31) the resonance comes out at 1965 MeV and in the exact case of (22) it comes out at 2051 MeV (table 3). No certain N-resonance is observed in this domain. Being close to the observed resonance domain around 2100-2200 MeV the state 1,$\,$3,$\,$7 may hide itself by mixing with ordinary N-states. Thus it could explain the new resonance N(2040) in  $m_{p\pi^{-}}$ invariant mass spectra from $J/\Psi \rightarrow p \pi^{-} \overline{n}$  [36]. The singlet 5,$\,$7,$\,$9 lies just above the free charm threshold for baryonic decay into $\Sigma^{+}_c (2455) {\rm{D}}^{-}$  at 4324 MeV (2454+1870 [27]) and should be visible (together with 3,5,11 at 4652 MeV) in neutron diffraction dissociation experiments like those in ref. [37]. It should also be visible in $ \gamma n \rightarrow p \pi^{-}$  photoproduction experiments like [38]. Other, lower lying singlets shown in table 3 might be visible in invariant mass $m_{p\pi^{-}}$ from  $B$ decay experiments like [39, 40]. 
\newline

\section*{7 Open questions}

The present work has three open ends: a very specific, momentarily halted end; a far horizon perspective and finally a more immediate and accessible opening. \newline
 1. It has not been possible to solve the general eq. (22) for the alleged electrically charged states. A suitable set of base functions analogues to the uncharged case (44) has not been found. \newline 
2. The introduction of the charge degree of freedom in (41) is supported by the fact that $u$(1) is the gauge group generating the electromagnetic interaction. This point is in need of elaboration. \newline
3. The projection mediated by the exterior derivative relates toroidal parameters to spacetime parameters and transfers structure from the group configuration space into transformation properties of the spacetime fields. However we also need to understand how the topology of the allospatial state transfers into angular distributions in spacetime partial waves. We expect $D$-functions to be involved. They are representations of the rotation group o(3) which happens to be doubly covered by su(2) of which there are three 'copies' in su(3), e.g. re\-presented by the three spins $U$, $V$ and $I$. We may then assume that the spin structure in allospace is excited via the generators of the rotation group in laboratory space. If we apply the so-called Clebsh-Gordan series [41] twice we can reduce a triple product of $D$-functions to a sum over $D$-functions presumably pertaining to the toroidal wavefunctions with specific angular momentum $j$
\begin{widetext}
\begin{gather} 
   D_{k_1m_1}^{j_1} D_{k_2m_2}^{j_2} D_{k_3m_3}^{j_3} =
   \sum_{j'} C(j_1 j_2j';k_1k_2) C(j_1 j_2j';m_1m_2) D^{j'}_{k_1+k_2,m_1 + m_2}
   D^{j_3}_{k_3m_3} 
 = \nonumber\\ \sum_{j'}\! C(j_1 j_2j';k_1k_2) C(j_1 j_2j';m_1m_2) \sum_j\!C(j' j_3 j;k_1\!+\! k_2, k_3)
   C(j' j_3 j;m_1\!+\! m_2, m_3)   
   D^j_{k_1 + k_2 + k_3, m_1 + m_2 + m_3}. \end{gather} \vspace{7mm} 
 \end{widetext}  
  
A systematic investigation of total spin and parity eigenstates based on $D$-functions like (37) seems an obvious task to undertake from here thus attacking point 3, but we would like to sort out the relation between the structure of the $D$-functions and the toroidal labeling of states, on which our present interpretation of charge multiplets rests. After all the energy eigenstates constructed from expansions on parametric function Slater determinants agree surprisingly well with the observed states as seen in fig. 3. Further the toroidal parton distributions traced out in appendix D and shown in fig. 7 compares rather well in shape with the proton valence quark distributions in figs. 16.4 of [42] and 6.14 of [43].

\begin{figure} [h]
\begin{center}
\includegraphics[width=0.45\textwidth]{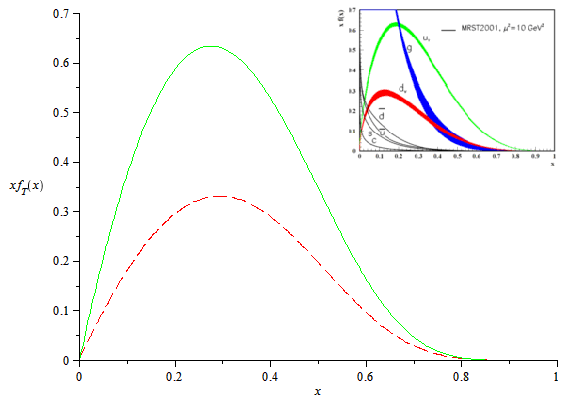}
\caption{Parton distribution functions for a first order approximation to an alleged proton ground state of (1). The distributions are traced out by specific toroidal generators $T_u=(\frac{2}{3},0,-1)$ and $T_d=(-\frac{1}{3},0,-1)$ via the exterior derivative (9), see appendix D. The curves generated by   $T_u$ (green solid) and $T_d$ (red dashed) compare both in peak value and relative integral content with the proton valence quark distributions $u_v$ and $d_v$ shown in the insert from [42].}
\label{fig7}
\end{center}
\end{figure}

\subsection*{7.2 On Interaction}

{\it{Ansatz}}

Baryons interact via the harmonic potential, their 'distance' being measured by the length $d(u,u')$  of the 'relative' geodetic where $u$ and $u'$ are the configuration variables of the respective baryons. This leads to two conjectures.

{\it{Conjecture 1}}

The deuteron should be the ground state of the following Schr\"{o}dinger equation
\begin{equation} 
    \frac{\hbar c}{a}\left[ - \frac{1}{2} \Delta_u - \frac{1}{2} \Delta_{u'} +  \frac{1}{2}d^2(u,u')
    \right] \Psi(u,u') = E \Psi(u,u').
\end{equation}

{\it{Conjecture 2}}

The isotopic landscape should in principle be derivable from a Hamiltonian something like
\begin{align} 
   {\rm{H}}_A =  {\rm{H}}_{Z+N} = \hspace{4.5cm} \nonumber \\ \frac{\hbar c}{a} \left[ \sum^{Z,N}_{i,j} - \frac{1}{2}( \Delta_i + \Delta_j) + \sum^{A(A-1)/2}_{k<l}  \frac{1}{2} d^2(u_k,u_l) \right]\!. \end{align}
Conjecture 1 should be tractable, at least numerically, by exploiting the translational invariance of the distance measure. Here one would rewrite
\begin{equation} 
   d(u,u')=d(e,u^{-1}u') = d(e,u^{\dagger}u'),
\end{equation}
 where in the last expression we have used the unitarity of $u$. In that way we end up with a system of six differential equations in the two sets of eigenangles. This set is coupled by the potential term
\begin{equation} 
   \frac{1}{2} d^2(e,u^{\dagger}u') = \frac{1}{2} \sum ^3_{j=1}(-\theta_j + \theta'_j)^2 
   =\frac{1}{2} \sum ^3_{j=1}(\theta'_j - \theta_j)^2.
\end{equation}
Conjecture 2 may be tested by studying the decay of tritium to helium-3. According to our allospatial hypothesis this decay involves period doublings analogous to the decay of the neutron and thus in principle one should be able to calculate the relative mass difference of  $^3_1$H and $^3_2$He  without knowing the actual mechanism behind the creation of the electron and its antineutrino.

\section*{8 Conclusions}

We have presented a radical approach to the confinement problem by shifting the dynamics of strong interaction baryon spectroscopy to a compact configuration space. We have developed a hamiltonian framework in which to formulate this dynamics and suggested how the dynamics communicate with spacetime kinematics. We have called it the allospatial hypothesis.
The model has no fitting parameters, except the scale $\Lambda$  which turns out to be 210 MeV corresponding to a length scale $a \approx 1$ fm.\linebreak The number of resonances and their grouping in the N and $\Delta$ sectors are reproduced well by eigenstates of a Schr\"{o}dinger equation on u(3), called allospace. A quite accurate prediction of the relative neutron to proton mass shift 0.138 \% follows from approximate solutions to the same Schr\"{o}dinger equation. A projection of states to space is given via the exterior derivative. This projection has shown to yield parton distribution functions that compare rather well with those of the proton valence quark distributions already in a first order approximation. A kinematic parametrization for the projection gives a natural transition between a confinement domain where the dynamics unfolds in the global group space and an asymptotic free domain where the algebra approximates the group. A promising ratio between the $\Delta$(1232) and N(939) masses has been calculated based on specific $D$-functions. We expect the allospatial energy eigenstate spectrum to project into partial wave amplitude resonances of specific spin and parity via expansions on specific combinations of $D$-functions. Singlet neutral flavour resonances are predicted above the free charm threshold of $\Sigma^{+}_c (2455)D^{-}$. The presence of such singlet states distinguishes experimentally the present model from the standard quark model as does the prediction of the neutron to proton mass splitting.

\section*{Acknowledgments}

We thank Torben Amtrup, Jeppe Dyre, Jakob Bohr, Mogens Stibius Jensen, Mads Hammerich, Povl Holm, Hans Bruun Nielsen, Holger Bech Nielsen, Norbert Kaiser, Vladimir B.\ Kopeliovich, Sven Bjørnholm and Bo-Sture Skagerstam for key comments. And we thank Erik Both for the meticulous transformation of our manuscript into \LaTeX. We have also had help from Geoffrey C. Oades, Abel Miranda, Poul Werner Nielsen, Jens Hugger, Bo Gervang, Ivan \v{S}tich, Karen Ter-Martirosyan, Yurii Makeenko, Dmitri Boulatov, Jaime Vilate, Pedro Bucido, Manfried Faber, A. Di Giacomo, Gestur Olafsson, M. Blazek, P. Filip, S. Olejník, M. Nagy, A. Nogova, Peter Presnajder, Vlado Cerný, Juraj Bohácik, Roman Lietava, Miroslava Smr\v{c}inová, 
Hans Plesner Jacobsen, Vracheslav P. Spiridonov, Norbert Kaiser, Poul Olesen, Ben Mottelson, Andreas Wirzba, Niels Kjær Nielsen, Bo-Sture Skagerstam, Per Salomonson, Palo Vaslo, Lissi Regin, Berit Bjørnow, Elsebeth Obbekjær Petersen, Hans Madsbøll, Karl Moesgen, Victor F. Weisskopf, Jørgen Kalckar, Leo Bresson, Knud Fjeldsted, Jens Bak, Mogens Hansen, Jane Hvolbæk Nielsen, Jens Lyng Petersen, Peter Hansen, Jørn Dines Hansen.

\appendix
\section{Projection to space}

For each element  $u\in u(3)$  we have a corresponding left translation $l_u$  on  $v \in u(3)$
\begin{equation} 
   l_u(v)\equiv uv,
\end{equation}
and for any left-invariant vector field  X we have [44]
\begin{equation} 
   \mbox{X}_{uv} =d(l_u)_v(\mbox{X}_v).
\end{equation}
In particular we have for our toroidal coordinate fields when comparing with (4)
\begin{equation} 
   \partial_j \!\mid_{u\cdot e} = d(l_u)_e(\partial_j \!\mid_e) = u \partial_j \!\mid_e\!.
\end{equation}
Thus the exterior derivative $d$  acts as the identity on left translations at the origo $e$. We now expand the exterior derivative of the measure scaled wave function $\Phi = J \psi$  on the torus forms (5), i.e.  
\begin{equation} 
   d \Phi= \psi_j d \theta_j,
\end{equation}
where the coefficients are the local partial derivatives [45]
  \begin{equation} 
    \psi_j(u)\equiv d\Phi_u (\partial_j), \quad j=1,2,3.
\end{equation}
For the coefficients we have by left invariance (A3) 
\begin{gather} 
   \psi_j(u)=d\Phi_u(\partial_j)= \partial_j \!\!\mid_u[\Phi] =u\partial_j \!\!\mid_e[\Phi] 
   \nonumber \\ = ud \Phi_e(\partial_j) = u \psi_j(e). \hspace{1cm} \end{gather}
The sum of the torus form partial derivative coefficients will inherit the left invariance
\begin{gather} 
   \overline{\psi}(u) \equiv \psi_1(u) + \psi_2(u) + \psi_3(u)  \nonumber \\ =
 u (\psi_1(e) + \psi_2(e) + \psi_3(e)) = u \overline{\psi} (e). \end{gather}
Now in particular $\psi_j(e)=d\Phi_e(\partial_j)=\partial\Phi/\partial\theta_j$ belongs to the tangent space $TM_e$ of the maximal torus $M$ at $e$ and therefore so does their sum $\overline{\psi}(e)$ 
as in general $\psi_j(u)\in TM_u$.\linebreak The set of generators  $\{ i T_j \}$ are the coordinate field generators $\partial_j$ which also constitute an induced base from parameter space
\begin{equation} 
  \partial_j \!\! \mid_u = \frac{\partial}{\partial\theta_j} \!\! \mid_u = d(exp)_{exp^{-1}(u)}(\vec{c_j}),
\end{equation}
where $\{\vec{c}_j \}$ is a set of base vectors for the parameter space for the torus, see fig. 9, appendix E. In our interpretation we identify $\{\vec{c}_j \}$ with a base for a fundamental representation space for the colour algebra SU(3). We may thus introduce complex valued components $\tilde{\psi_j}$ for the colour vector $\overline{\psi}$ and write
\begin{equation} 
   \overline{\psi} = \tilde{\psi}_1 \vec{c}_1 + \tilde{\psi}_2 \vec{c}_2 + \tilde{\psi}_3 \vec{c}_3 =
\begin{Bmatrix}  \tilde{\psi}_1 \\ \tilde{\psi}_2 \\ \tilde{\psi}_3 \end{Bmatrix}\!.
\end{equation}
In the above representation $u$ will be represented by a $3\times 3$ matrix $U$\!. For rotations under $V\in SU(3)$ we then have
\begin{equation} 
   \vec{c}_j \rightarrow \vec{c'}_j = V\vec{c}_j 
\end{equation}
and
\begin{equation} 
   U \rightarrow U'=VUV^{-1}.
\end{equation}
From (A7), (A10) and (A11) we can derive the transformation property of $\overline{\psi}(u)$.
\begin{gather} 
   \overline{\psi}(u) = U \overline{\psi}(e) \rightarrow \overline{\psi}(u') ' = U' \overline{\psi}(e) ' =
   \nonumber \\ VUV^{-1}V \overline{\psi}(e) = VU \overline{\psi}(e) = V \overline{\psi} (u),
\end{gather}
which shows that the differential coefficient vector $\overline{\psi}$ transforms as a colour vector in the fundamental re\-presentation under SU(3) rotations. In other words left invariance in group space projects out as SU(3) rotation in projection space. We thus interpret $\overline{\psi}$  in (A7) as a quark field with three colour components which may be projected on a specific base like in (A9). 

Likewise the gluon fields may be seen as resulting from a projection on adjoint representation spaces of an expansion of the momentum form corresponding to the full set of eight generators $\lambda_k$  needed to parametrize the general group element $u=e^{i\alpha_k\lambda_k}$  -- separating out a phase factor [46]. Thus for each generator $T_a$  we have left invariant vector fields $\partial_a$  defined as
\begin{equation} 
   \partial_a = \frac{\partial}{\partial \omega} e^{i\alpha_k T_k} e^{i \omega T_a} \!\! \mid_{\omega=0}
   = u i T_a,
\end{equation}
where $T_a=-i\partial/\partial \alpha_a = - i \partial_a\! \! \mid_e$. We now choose the set $\{T_a\}$  as a base for the adjoint representation. This base transforms under $V \in SU(3)$  like
\begin{equation} 
   T_a' = VT_aV^{-1}.
\end{equation}
Analogous to (A4) we expand the external derivative $d\Phi$  on forms related to the left invariant vector fields $\partial_a$   to get the adjoint projection field
\begin{equation} 
   \overline{A}(u) = \sum_a d \Phi_u(\partial_a).
\end{equation}
We want to show that $\overline{A}$  transforms according to the adjoint representation. First we have the equivalent of (A7)
\begin{gather} 
   \overline{A}(u) = \sum_a d \Phi_u(\partial_a)
      =\sum_a \partial_a \!\! \mid_u [\Phi]
      \nonumber \\= \sum_a u \partial_a \!\! \mid_e [\Phi]
      =u \sum_a d\Phi_e(\partial_a)=u\overline{A}(e).
\end{gather}
Here we understand in analogy with (A9) that
\begin{equation}
	\overline{A}(e)= \tilde{A}_a(e)T_a,
\end{equation}
where again $\tilde{A}_a$ are complex valued components.
We may then proceed to show the adjoint transformation property of $\overline{A}$
\begin{gather}
	\overline{A}(u) \rightarrow \overline{A}(u') ' = U'\overline{A}(e) ' = U'\tilde{A_a}(e)T_a'
	\nonumber \\ =VUV^{-1} \tilde{A_a}(e)VT_aV^{-1} =VUV^{-1}V \tilde{A_a}(e)T_aV^{-1} 
	\nonumber \\
	= VU \overline{A}(e)V^{-1} = V \overline{A}(u)V^{-1},
\end{gather}
which corresponds to the gauge group rotation transformation property of the gluon fields $B$ [47]
\begin{equation}
	B_{\mu} ' = VB_{\mu}V^{-1}+\frac{i}{g}(\partial_{\mu}V)V^{-1},
\end{equation}
where
\begin{equation}
	V = e^{-i \alpha_a(x) T_a}.
\end{equation}
We note that as space time fields the gauge fields also acquire a term representing the variation along spacetime translations as represented by the second term in (A19). Note further that translational invariance in group space corresponds to an SU(3) rotational invariance in representation space and thereby the translational invariance of the interaction potential (3) in group space through the projections (6) and (9) reflects the gauge invariance of the fields in laboratory space.

\subsection{Kinematics}

Using (6) and (7) and suppressing the centrifugal and curvature terms like in (31), the projected Schr\"{o}dinger equation reads

\begin{equation} 
   \left[- \frac{1}{2} a^2 \frac{\partial}{\partial x_j} \frac{\partial}{\partial x_j}
   + \frac{1}{2} \frac{1}{a^2} d^2({\mbox{\bf{x}}})\right] R( {\mbox{\bf{x}}}/a) 
   = \frac{E a}{\hbar c} R( {\mbox{\bf{x}}}/a) .
\end{equation}

 In the high energy limit, taking $E\rightarrow \infty$  to be the la\-boratory energy and letting $a\rightarrow 0$  while the allospatial eigenvalue E = $\frac{E a}{\hbar c}$  remains fixed, we see that the projected potential blows up and the kinetic term cools down to restrict the support of $R$ to the neighbourhood of origo where the algebra approximates the group, see fig. 1. Compactification loosens up. This is a sign of asymptotic freedom. In the other extreme taking $E\rightarrow E_0 = m_0c^2$  we will have $a$ increasing, thereby lowering the potential such that the wavefunction spreads out in all of the group. Adjusting $E$ in the experimental set-up automatically determines whether one sees the deglobalized (asymptotically free) or the globalized (confining) characteristics prevail.

   From the projection (6) and (7) to laboratory space we recognize the toroidal generators as momentum operators. Thus when experimental production of resonances is of concern we see from space: {\it{ The impact momentum generates the (abelian) maximal torus of the u(3) allospace. The momentum operators act as introtangling generators.}} When decay, asymptotic freedom, fragmentation and confinement is of concern we see from allospace: {\it{The quark and gluon fields are projections of the vector fields induced by the momentum form}} $d \Phi$.

\section{The spectrum of ${\mbox{\bf{K}}}^2$  and ${\mbox{\bf{M}}}^2$}

In the allospatial interpretation the presence of the components of {\bf{K}} $= (K_1,K_2,K_3) $  and \linebreak {\bf{M}}~$= (M_1,M_2,M_3) $  in the Laplacian opens for the inclusion of spin and non-neutral flavour. It can be shown [14] that the components commute with the Laplacian as they should since the Laplacian is a Casimir operator and therefore the generators of the group commutes with it. They also commute with the geodetic potential, so
\begin{equation} 
  [K_k, \mbox{H}] =  [M_k, \mbox{H}] =0,
\end{equation}						
where H is the Hamiltonian in (1). Further
\begin{equation} 
  [K_k,\mbox{\bf{K}}^2] =  [K_k,\mbox{\bf{M}}^2] = [M_k,\mbox{\bf{M}}^2] 
= [M_k,\mbox{\bf{K}}^2]    = 0.
\end{equation}

Thus we may choose {\bf{K}}$^2$, $K_3$, {\bf{M}}$^2$ as a set of mutually commuting generators which commute with the Hamiltonian H. To solve the general problem (1) or (10) we need to specify the spectrum of the involved operators. The well-known eigenvalues of {\bf{K}}$^2$ and $K_3$ follow [48] from the commutation relations (15). Here we choose to interpret {\bf{K}} as an interior angular momentum operator and allow for half-integer eigenvalues. Instead of choosing eigenvalues of $K_3$ we may choose $I_3$, the isospin 3-component.

To determine the spectrum for {\bf{M}}$^2$ we introduce a canonical body fixed 'coordinate' representation [49]
\begin{align} 
   K_1& = a \theta_2p_3 - a \theta_3 p_2 = \hbar \lambda_7 \nonumber \\ 
     K_2& = a \theta_1p_3 - a \theta_3 p_1 = \hbar \lambda_5 \nonumber \\
     K_3& = a \theta_1p_2 - a \theta_2 p_1 = \hbar \lambda_2. 
\end{align}
The remaining Gell-Mann generators are traditionally collected into a quadrupole moment tensor {\bf{Q}}, but we need to distinguish between the two diagonal components
\begin{gather} 
  Q_0/ \hbar  =\hspace{3cm} \nonumber \\
     \frac{1}{2\sqrt{3}} ( \theta^2_1 + \theta^2_2- 2\theta^2_3) 
   + \frac{1}{2\sqrt{3}} \frac{a^2}{{\hbar}^2}(p^2_1 + p^2_2- 2p^2_3) =\lambda_8, \nonumber \\
  Q_3/ \hbar  =  \frac{1}{2} ( \theta_1^2 - \theta^2_2) + \frac{1}{2}  \frac{a^2}{{\hbar}^2}
   (p_1^2 - p^2_2) = \lambda_3
\end{gather}
and the three off-diagonal components which we have collected into {\bf{M}}
\begin{align} 
  M_3/\hbar &= \theta_1 \theta_2 +  \frac{a^2}{{\hbar}^2} p_1 p_2 = \lambda_1 \nonumber \\
  M_2/\hbar &= \theta_3 \theta_1 +  \frac{a^2}{{\hbar}^2} p_3 p_1 = \lambda_4 \nonumber \\
  M_1/\hbar &= \theta_2 \theta_3 +  \frac{a^2}{{\hbar}^2} p_2 p_3 = \lambda_6 .
\end{align}
{\bf{M}} is a kind of Runge-Lenz 'vector' [50] of our problem. This is felt already in its commutation relations (15). We shall see in the end (B13) that conservation of {\bf{M}}$^2$ corresponds to conservation of particular combinations of hypercharge and isospin.

For the spectrum in projection space we calculate the SU(3) Casimir operator using the commutation relations (8)
\begin{equation} 
   C_1 = \frac{1}{\hbar} ( \mbox{\bf{K}}^2 + \mbox{\bf{M}}^2 + Q^2_0 + Q^2_3) 
  = -3 + \frac{1}{3 \Lambda^2} (2\mbox{H}_e)^2,
\end{equation}
where the Hamiltonian H$_e$  of the euclidean harmonic oscillator is given by
\begin{align} 
   2\mbox{H}_e = \frac{ca}{\hbar} \mbox{\bf{p}}^2 + \frac{\hbar c}{a}  {\mbox{\boldmath{$\theta$}}}^2, \hspace{2cm} \nonumber \\
   \mbox{with} \quad \mbox{\bf{p}} = (p_1, p_2, p_3), \, \, \,\, {\mbox{\boldmath{$\theta$}}} = (\theta_1, \theta_2, \theta_3).
\end{align}

We define canonical annihilation and creation operators
\begin{align} 
&a_j = \frac{1}{\sqrt{2}} \left( \frac{a}{\hbar} p_j - i \theta_j \right) \nonumber \\
& a^{+}_j = \frac{1}{\sqrt{2}} \left( \frac{a}{\hbar} p_j + i \theta_j \right) \nonumber \\ 
&[a_i , a^{+}_j ] = \delta_{ij}, \quad [a_i , a_j ] = [a_i^{+} , a_j^{+} ] =0,
\end{align} 
and now want to fix the interpretation of the two diagonal operators $Q_0$ and $Q_3$. We find
\begin{align} 
   Y/\hbar &\equiv \frac{Q_0/ \hbar}{\sqrt{3}} = \frac{N}{3} - a_3^{+} a_3\nonumber \\
   2I_3/ \hbar & \equiv Q_3/ \hbar = a_1^{+} a_1 - a_2^{+} a_2 \nonumber \\
   Q_2/ \hbar & = a_1^{+} a_1 - a_3^{+} a_3,
\end{align}
where the number operator
\begin{equation} 
    N \equiv \sum^3_{j=1} a_j^{+} a_j
\end{equation}				
and
\begin{equation} 
   Q_2/ \hbar \equiv \frac{\sqrt{3} Q_0 + Q_3}{2 \hbar} = \frac{1}{2} ( \theta_1^2 - \theta_3^2) +
   \frac{1}{2} \frac{a^2}{\hbar^2}(p_1^2 - p_3^2).
\end{equation}
From (B9) we get
\begin{equation} 
    Y = \frac{1}{3} (2Q_2 - 2 I_3).
\end{equation}				

Provided we can interpret $Q_2$ as a charge operator this is the well-known eight-fold way relation of Gell-Mann and Ne'eman between charge, hypercharge and isospin [51, 52]. Inserting (B9) in (B6) and rearranging we get
\begin{equation} 
   \mbox{\bf{M}}^2 = \frac{4}{3}
\hbar^2 \left( \frac{\mbox{H}_e}{\Lambda} \right)^2 - \mbox{\bf{K}}^2 - 3 \hbar^2 - 3 Y^2 - 4I_3^2.
\end{equation}				

The spectrum of the three-dimensional euclidean isotropic harmonic oscillator Hamiltonian (B7) is well-known [53, 54]. If we assume the standard interpretations in (B12) with $Q_2$ as a charge operator we have a relation among quantum numbers ($y_{\rm{quantum\, \, number}}\, \sim \, 3 Y_{\rm{operator}} /\hbar$) from which to determine the spectrum of  $\mbox{\bf{M}}^2/ \hbar^2$, namely
\begin{align} 
  M^2 = \frac{4}{3}(n+\frac{3}{2})^2 - K(K+1) - 3  -\frac{1}{3}y^2 - 4i^2_3, \nonumber \\
  n = n_1 +n_2 +n_3, \quad n_i = 0,1,2,3, \dots \end{align}
Now {\bf{M}} is hermittean and therefore $M^2$ must be non-negative. With $K=\frac{1}{2}, \, \, y=1,  \,\, i_3 = \pm \frac{1}{2}$ as for the nucleon, the lowest possible value for $n$ is 1 (where $M^2 = 13/4$). For the combination  $y=1, \, \, i_3= \pm \frac{1}{2}$, with $n=1$  we have $K(K+1)+M^2=4$  and thus only two half integer values are possible for $K$, namely $K=\frac{1}{2}, \, \, M^2= \frac{13}{4}$  and $K=\frac{3}{2}, \, \, M^2= \frac{1}{4}$, whereas $K=\frac{5}{2}$  would demand an impossible negative $M^2$. If we keep  $y=1$  and ask for $i_3= \pm \frac{3}{2}$  the minimum value of $n$ is 2 and again we get $K(K+1)+M^2=4$ allowing for the two above mentioned values of $K$. Note that $n=3$  does not allow  $i_3= \pm \frac{5}{2}$    if we keep $y=1$.

The relation in (B13) can be cast into an Okubo-form by choosing a different set of mutually commuting operators. We want to replace the three-component of isospin by isospin itself. This is possible because
 \begin{equation} 
   I^2 = I_1^2 + I_2^2 + I_3^2= \frac{1}{4} ( K_3^2 + M_3^2) +I_3^2
\end{equation}				
and $[K_3^2 +M_3^2,I^2] =0$. Thus by rearranging (B15) and (B13) we get
\begin{equation} 
     \mbox{\bf{K}}^2 + \mbox{\bf{M}}^2 = \frac{4}{3} \hbar^2  \left( \frac{H_e}{\Lambda} \right)^2
   - 3 \hbar^2 + (K_3^2 + M_3^2) - 3Y^3 - 4I^2.
\end{equation}		
From (B16) we get the following relation among quantum numbers
\begin{equation} 
   K(K+1) +M^2 \!=\! \frac{4}{3}(n+ \frac{3}{2})^2 - 3 + (k_3^2 + m_3^2) - \frac{1}{3} y^2 - 4 i(i+1).
\end{equation}
Here  $(k_3^2 + m_3^2)$ should be considered a single quantum number. For a given value of $(k_3^2 + m_3^2)$ we may group the spectrum in (B17) according to  $n+y =$ constant and get the Okubo structure (26) for the nominator in the centrifugal potential as described in the main text.

\section{Rayleigh-Ritz solution}

We want to find the eigenvalues E of the following equation
\begin{equation}    
 [ -\Delta_e+V]R(\theta_1,\theta_2,\theta_3) = 2 \mbox{E}R(\theta_1,\theta_2,\theta_3),
\end{equation}
which is the full eq. (22) in the main text.

In the Rayleigh-Ritz method [55] one expands the eigenfunction on an orthogonal set of base functions with a set of expansion coefficients, multiply the equation by this expansion, integrates over the entire variable volume and end up with a matrix problem in the expansion coefficients from which a set of eigenvalues can be got. Thus with the approximation
\begin{equation}    
R_N = \sum^N_{l=1} a_l f_l
\end{equation}
we have the integral equation
\begin{gather}    
  \int^{\pi}_{-\pi}\int^{\pi}_{-\pi}\int^{\pi}_{-\pi} R_N \cdot ( -\Delta_e+V) 
   R_N d\theta_1 d\theta_2d\theta_3 \nonumber \\
  =\int^{\pi}_{-\pi}\int^{\pi}_{-\pi}\int^{\pi}_{-\pi} R_N \cdot
  2 \mbox{E} R_N d\theta_1 d\theta_2 d\theta_3.\end{gather}
The counting variable $l$ in (C2) is a suitable ordering of the set of tripples $p$, $q$, $r$ in (44) or $l$, $m$, $n$ in (33) such that we expand on an orthogonal set. The eq. (C3) can be interpreted as a vector eigenvalue problem, where {\bf{a}} is a vector, whose elements are the expansion coefficients $a_l$. Thus (C3) is equivalent to the eigenvalue problem
\begin{equation}    
  \mbox{\bf{a}}^T \mbox{\bf{Ga}} = 2 \mbox{E} \mbox{\bf{a}}^T \mbox{\bf{Fa}},
\end{equation}
where the matrix elements of {\bf{G}} and {\bf{F}} are given by
\begin{equation}    
    G_{lm} \equiv \int^{\pi}_{-\pi}\int^{\pi}_{-\pi}\int^{\pi}_{-\pi} f_l \cdot  ( -\Delta_e+V)  f_m 
   d\theta_1 d\theta_2 d\theta_3
\end{equation}
and
 \begin{equation}    
    F_{lm} \equiv \int^{\pi}_{-\pi}\int^{\pi}_{-\pi}\int^{\pi}_{-\pi} f_l f_m 
   d\theta_1 d\theta_2 d\theta_3. 
\end{equation}				

When the set of expansion functions is orthogonal, (C4) implies
\begin{equation}    
  {\mbox{\bf{Ga}}}= {\mbox{2E{\bf{Fa}}}},
\end{equation}
from which we get a spectrum of $N$ eigenvalues determined as the set of components of a vector {\bf{E}} generated from the eigenvalues of the matrix ${\bf{F}}^{-1}{\bf{G}}$, i.e.
\begin{equation}    
  {\mbox{\bf{E}}}= \frac{1}{2} {\mbox{eig}} ({\mbox{\bf{F}}^{-1}} {\mbox{\bf{G}}}).
\end{equation}

The lowest lying eigenvalues will be better and better determined for increasing values of $N$ in (C2). For the base (44) the integrals (C5) and (C6) can be solved ana\-lytically, and as (44) is an educated guess based on the solution of the separable problem (31) it improves the convergence in $N$ for the general problem in (C1) and (C7).

The exact expressions to be used in constructing {\bf{G}} and {\bf{F}} are given below for the base (44). For $r > p,\, u > s$ and $q,\,t \ge 1$ we have the following orthogonality relations
\begin{gather}    
   <f_{pqr}\! \mid \! f_{stu}\! > \equiv \hspace{4cm} \nonumber \\
\int^{\pi}_{-\pi}\int^{\pi}_{-\pi}\int^{\pi}_{-\pi}
   f_{pqr}(\theta_1,\theta_2,\theta_3) \cdot f_{stu} (\theta_1,\theta_2,\theta_3)
   d\theta_1 d\theta_2 d\theta_3 \nonumber \\
 =6 \pi^3 \delta_{ps}\delta_{qt}\delta_{ru}.
 \end{gather}
Here for a convenient notation we have generalized the Kronecker delta
\begin{equation}   
     \delta_{ij} = 
 \begin{cases} 
  1  \, \, \, & \text{for} \, \, \,  i=j \, \wedge \, i \ne 0  \\
  2  & \text{for} \, \, \,  i=j \, \wedge \, i = 0 \\
  0   & \text{for}\, \, \,  i \ne j
 \end{cases}.   
\end{equation}
The Laplacian yields
\begin{gather}    
  <f_{pqr} \! \mid \! \frac{\partial^2}{\partial \theta^2_1} + \frac{\partial^2}{\partial \theta^2_2} +   
 \frac{\partial^2}{\partial \theta^2_3} \! \mid \! f_{stu}>\nonumber \\
  = ( - p^2 - q^2 - r^2) \cdot 
  6 \pi^3 \delta_{ps}\delta_{qt}\delta_{ru}.
\end{gather}
The matrix elements for the geodetic potential couples the individual base functions and follows from a more lengthy calculation below yielding the following expression
\begin{eqnarray}    
 &<\! f_{pqr} \! \mid \! \theta_1^2 +\theta_2^2 +\theta_3^2 \! \mid \! f_{stu}\!> =  \nonumber \\ 
 & \nonumber \\
&\text{for }p,q,r,s,t,u >0: \nonumber \\
&\delta_{ps}\delta_{qt}\delta_{ru} \cdot 6 \pi^3 \left(\pi^2 + \frac{1}{2p^2}- 
\frac{1}{2q^2}+\frac{1}{2r^2} \right)\nonumber  \\
&+(1-\delta_{ps})\delta_{qt} \delta_{ru} \cdot 6 \pi^3 \cdot 4 \frac{p^2+s^2}{(p^2-s^2)^2} \cdot(-1)^{p+s} \nonumber\\
&+(1-\delta_{qt})\delta_{ps} \delta_{ru} \cdot 6 \pi^3 \cdot 4 \frac{2qt}{(q^2-t^2)^2} \cdot(-1)^{q+t} \nonumber\\
&+(1-\delta_{ru})\delta_{ps} \delta_{qt} \cdot 6 \pi^3 \cdot 4 \frac{r^2+u^2}{(r^2-u^2)^2} \cdot(-1)^{r+u} \nonumber\\
&-(1-\delta_{pu})\delta_{qt} \delta_{rs} \cdot 6 \pi^3 \cdot 4 \frac{p^2+u^2}{(p^2-u^2)^2} \cdot(-1)^{p+u} \nonumber\\
&-(1-\delta_{rs})\delta_{pu} \delta_{qt} \cdot 6 \pi^3 \cdot 4 \frac{r^2+s^2}{(r^2-s^2)^2} \cdot(-1)^{r+s} \nonumber\\
& \nonumber\\
&\text{for } p=0 \, \wedge \, s \ne 0 \, \wedge \, u> s: \nonumber \\
& 24 \pi^3 \left[ \delta_{qt} \delta_{ru} \frac{1}{s^2} (-1)^s - \delta_{qt} \delta_{rs} \frac{1}{u^2}(-1)^{u} \right] \nonumber\\
& \nonumber\\
&\text{for } p=0 \, \wedge \, s=0: \nonumber \\
& \delta_{qt} \delta_{ru} \cdot 6 \pi^3 ( 2 \pi^2 - \frac{1}{q^2} + \frac{1}{r^2} ) \nonumber \\
& +(1-\delta_{qt} ) \delta_{ru} \cdot 48 \pi^3 \frac{2 q t}{(q^2-t^2)^2}\cdot (-1)^{q+t}  \nonumber \\
& +(1-\delta_{ru} ) \delta_{qt} \cdot 48 \pi^3 \frac{r^2 + u^2}{(r^2-u^2)^2}\cdot (-1)^{r+u}. 
\end{eqnarray}

Finally the integrals needed for the matrix elements of the centrifugal potential can be solved by a change of variables.  Exploiting the periodicity of the trigonometric functions the domain of integration can be selected to suit the new set of variables, see fig. 8 and the section below on elementary integrals.

The result is
\begin{widetext}
\begin{align}
 &<\! f_{pqr} \! \mid \frac{1}{\sin^2 \frac{1}{2} (\theta_1 - \theta_2)} + \frac{1}{\sin^2 \frac{1}{2} (\theta_2 - \theta_3)} + \frac{1}{\sin^2 \frac{1}{2} (\theta_3 - \theta_1)} \mid f_{stu} > \nonumber \\
&= 3 <\! f_{pqr} \! \mid \frac{1}{\sin^2 \frac{1}{2} (\theta_1 - \theta_2)}  \mid f_{stu} > \nonumber \\
&= 3 \pi^3[ \delta_{ps}(\delta_{r-q,u-t}nn(r+q,u+t) -  \delta_{r-q,u+t}nn(r+q,u-t))\nonumber \\
&+  \delta_{ps}(-\delta_{r+q,u-t}nn(r-q,u+t) +  \delta_{r+q,u+t}nn(r-q,u-t))\nonumber \\
&+  \delta_{pu}(\delta_{r-q,s+t}nn(r+q,s-t) -  \delta_{r-q,s-t}nn(r+q,s+t))\nonumber \\
&+ \delta_{pu}(-\delta_{r+q,s+t}nn(r-q,s-t) +  \delta_{r+q,s-t}nn(r-q,s+t))\nonumber \\
&+   \delta_{qt}(\delta\delta_{p+r,s+u}nn(r-p,u-s) + \delta  \delta_{p+r,u-s}nn(r-p,u+s))\nonumber \\
&+   \delta_{qt}(\delta\delta_{r-p,s+u}nn(p+r,u-s) + \delta  \delta_{r-p,u-s}nn(p+r,u+s))\nonumber \\
&+  \delta_{rs}(\delta_{p+q,u-t}nn(p-q,u+t) -  \delta_{p+q,u+t}nn(p-q,u-t))\nonumber \\
&+   \delta_{rs}(-\delta_{p-q,u-t}nn(p+q,u+t) +  \delta_{p-q,u+t}nn(p+q,u-t))\nonumber \\
&+  \delta_{ru}(\delta_{p+q,s+t}nn(p-q,s-t) -  \delta_{p+q,s-t}nn(p-q,s+t))\nonumber \\
&+  \delta_{ru}(-\delta_{p-q,s+t}nn(p+q,s-t) +  \delta_{p-q,s-t}nn(p+q,s+t))],
\end{align}
\end{widetext}
where two more shorthand notations have been introduced
\setcounter{equation}{13}
\begin{equation}   
     \delta \delta_{ij} = 
 \begin{cases} 
  1  \, \, \, & \text{for} \, \, \,  i=j \, \wedge \, i \ne 0  \\
  -1  & \text{for} \, \, \,  i=-j \, \wedge \, i \ne 0 \\
  0   & \text{otherwise}
 \end{cases}   
\end{equation}
 and
\begin{equation}   
     nn(i,j) = 
 \begin{cases} 
  \mid i+j \mid - \mid i-j \mid   \, \, \, & \text{for} \, \, \,  i+j \equiv 0 \mod 2  \\
  0  & \text{otherwise.} 
 \end{cases}   
\end{equation}
The factor $nn$ originates from the following rule [56]
\begin{gather}   
     \int^{\pi}_{-\pi} \frac{\sin mx\cdot \sin nx}{\sin^2 x}dx = \nonumber \\
 \begin{cases} 
  (\mid m+n \mid - \mid m-n \mid ) \pi  \, \, \, & \text{for} \, \, \,  m-n \equiv 0 \mod 2  \\
  0  & \text{for} \, \, \,  m-n \equiv 1 \mod 2.
 \end{cases}   
\end{gather}
The integrals in (C16) pop up after the aforementioned change of variables which exploits the following trigonometric relations
\begin{gather}
\cos px \cos ry - \cos rx \cos py = \nonumber \\  \sin nu \sin mt + \sin mu \sin nt,  \nonumber \\ 
u =\frac{x+y}{2}, \,\, t = \frac{x-y}{2},\,\, n = r+p,\,\, m = r-p,
\end{gather}
and
\begin{gather}
\cos px \sin qy - \sin qx \cos py = \nonumber \\ \cos nu \sin mt - \cos mu \sin nt, \quad n=p+q, \,\,m=p-q. 
\end{gather}

\begin{figure}
\begin{center}
\includegraphics[width=0.4\textwidth]{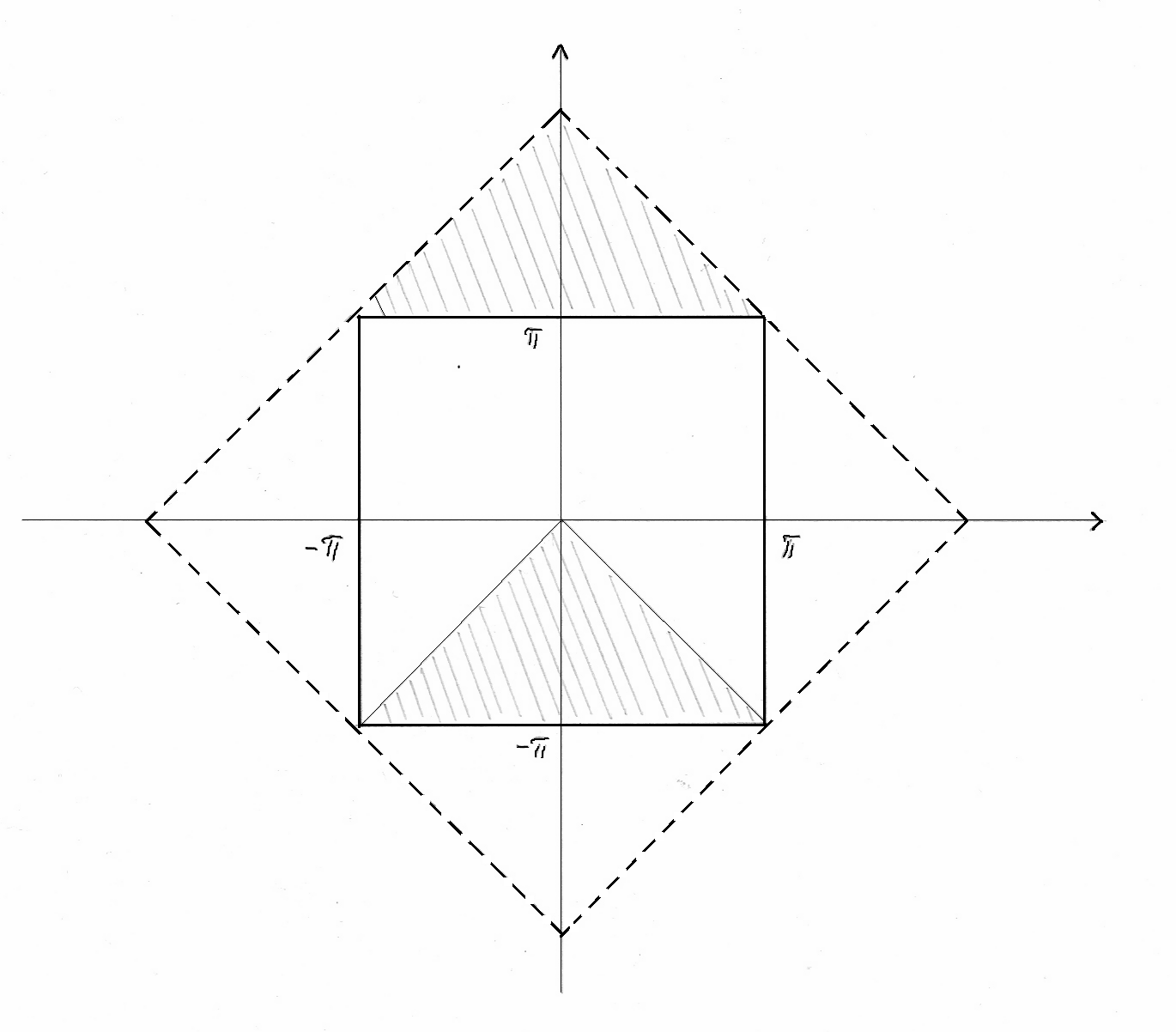}
\caption{A change of variables from the horizontal/vertical $(x,y)$ to a 45 degrees inclined system of coordinates $(u,t)=(\dfrac{x+y}{2},\dfrac{x-y}{2})$ needed in order to find the matrix elements of the centrifugal potential. The seemingly singular denominator in the centrifugal potential is then found to be integrable, see (C16) to (C18).  The domain of integration is expanded to suit the new set of variables. This is possible because of the periodicity of the trigonometric functions such that functional values on the hatched area outside the original domain of integration $[-\pi,\pi]\times[-\pi,\pi]$ are identical by parallel transport from the hatched area within that same area.}
\label{figC1}
\end{center}
\end{figure}

\subsection{Elementary integrals for matrix elements in the Rayleigh-Ritz method}

We solve here exemplar integrals for the trigonometric basis needed to prove the orthogonality relation (C9), the expectation value of the geodetic potential (C12) and the centrifugal potential (C13). First the orthogonality relation for   $p>0, \, q>0, \, r>p$  ($p=0$ is left for the reader). With a slight change in notation for the angular variables we seek the scalar product 
\begin{equation}   
   <f,g> \equiv \int^{\pi}_{-\pi}  \int^{\pi}_{-\pi}  \int^{\pi}_{-\pi}  fg dx_i dx_j dx_k
\end{equation}			
between base functions (44) like
\begin{gather}
f_{pqr}(x_i, x_j, x_k) = \epsilon_{ijk} \cos p x_i \sin q x_j \cos rx_k \quad {\text{and}}  \nonumber \\
g_{stu}(x_l, x_m, x_n) = \epsilon_{lmn} \cos p x_l \sin q x_m \cos rx_n. 
\end{gather}

The three-dimensional integral in (C19) factorizes into three one-dimensional integrals and using the orthogonality of the trigonometric functions on $[-\pi, \pi]$   we readily have
\begin{equation}
     <f_{pqr},g_{stu}> \, = ( \delta_{ps}  \delta_{ru}- \delta_{pu} \delta_{rs}) \delta_{qt} \cdot 6 \pi^3,
\end{equation}
which for $p>0, \, q>0, \,  r>p$ and $s>0, \, t>0, \, u>s$  reduces to (C9). To obtain the expectation value (C12) of the geodetic potential we use the same kind of factorization together with the following list of elementary integrals
\begin{gather} 
    \int^{\pi}_{-\pi} x^2 \cos px \cos sx \, ds = (-1)^{p+s} 4 \pi \frac{p^2 + s^2}{(p^2-s^2)^{2}}, \quad p \ne s \\
    \int^{\pi}_{-\pi} x^2 \cos sx \, dx= \frac{4 \pi}{s^2} (-1)^{s}, \quad p=0, \, s \ne 0 \hspace{17mm} \\
    \int^{\pi}_{-\pi} x^2 \cos^2 px \, dx = \frac{\pi^3}{3} + \frac{\pi}{2 p^2}, \quad p=s \hspace{25mm} \\ 
    \int^{\pi}_{-\pi} x^2 \sin^2qx \, dx = \frac{\pi^3}{3} - \frac{\pi}{2 q^2}, \quad q=t \hspace{26mm} \\
    \int^{\pi}_{-\pi} x^2 \sin qx \, \sin tx \, dx = \! (-1)^{q+t}\, 4 \pi \frac{2qt}{(q^2 - t^2)^{2}}, \hspace{2mm}
    | q | \ne | t |\\ 
    \int^{\pi}_{-\pi} x^2 \cos px \, \sin tx \, dx =0. \hspace{42mm}
\end{gather}

For the Laplacian and for the centrifugal potential we make another slight change in notation for our angular variables and rewrite our base functions (44) as a sum of subdeterminants
\begin{gather}
 f_{pqr}(x,y,z) = 
\begin{vmatrix}
 \cos px & \cos py & \cos pz \\
\sin qx & \sin qy & \sin qz \\
 \cos rx & \cos ry & \cos rz   
\end{vmatrix} \hspace{5.9cm} \nonumber \\
= \cos pz 
\begin{vmatrix}
\sin qx & \sin qy\\
\cos rx & \cos ry
\end{vmatrix} \hspace{13mm} \nonumber \\
- \sin qz 
\begin{vmatrix}
\cos px & \cos py\\
\cos rx & \cos ry
\end{vmatrix} \hspace{12.8mm} \nonumber \\
+ \cos rz 
\begin{vmatrix}
\cos px & \cos py\\
\sin qx & \sin qy
\end{vmatrix}. \hspace{11.8mm}
\end{gather}
From (C28) we then get terms like
\begin{gather}
   \frac{\partial^2}{\partial z^2} f_{pqr} (x,y,z) =-  p^2 \cos pz
\begin{vmatrix}
\sin qx  \sin qy\\
\cos rx  \cos ry
\end{vmatrix}
\nonumber \\
 + q^2 \sin qz 
\begin{vmatrix}
\cos px  \cos py\\
\sin qx  \sin qy
\end{vmatrix} \nonumber \\
-r^2 \cos rz
 \begin{vmatrix}
\cos px  \cos py\\
\sin qx  \sin qy
\end{vmatrix}
\end{gather}
for the Laplacian. Again the integral for the expectation value factorizes into one-dimensional integrals where the orthogonality of the trigonometric functions can be exploited to obtain (C11).

For the centrifugal term the three-dimensional integral does not readily factorize. We need a change of variables which suits the mixing of the variables in the denominators. Due to the arbitrary labelling of our angles we have
\begin{widetext}
\begin{gather}
 < f_{pqr} \! \mid  \! \frac{1}{\sin^2 \frac{1}{2}(x-y)} +\frac{1}{\sin^2 \frac{1}{2}(y-z)} +
\frac{1}{\sin^2 \frac{1}{2}(z-x)} \! \mid \! f_{stu} >   \nonumber \\
 = 3  <f_{pqr} \! \mid  \! \frac{1}{\sin^2 \frac{1}{2}(x-y)} \! \mid \! f_{stu}>.
\end{gather}
With
\begin{equation}
  f_{pqr} f_{stu} = 
\begin{vmatrix}
    \cos px & \cos py & \cos pz \\
\sin qx & \sin qy & \sin qz \\
\cos rx & \cos ry & \cos rz 
\end{vmatrix}
\begin{vmatrix}
    \cos sx & \cos sy & \cos sz \\
\sin tx & \sin ty & \sin tz \\
\cos ux & \cos uy & \cos uz 
\end{vmatrix},
\end{equation}
we can use the subdeterminant expressions in (C28) to get e.g.\ a factor $3 \delta_{qt} \pi$  from the $z$-integration of the term involving the two sines while the product of the two corresponding subdeterminants is used for a shift of variables, see below. We have
\vspace*{-3mm}
 \begin{equation}
\begin{vmatrix}
   \cos px& \cos py\\
   \cos rx & \cos ry  
\end{vmatrix}
\begin{vmatrix}
   \cos sx& \cos sy\\
   \cos ux &\cos uy  \nonumber 
\end{vmatrix}\\
\end{equation}
\vspace*{-6mm}
 \begin{eqnarray}
&=& \!\!\! (\cos px \cos ry - \cos rx \cos py)(\cos sx \cos uy - \cos ux \cos sy) \hspace{3.2cm} \nonumber \\
&=& \frac{1}{2} \frac{1}{2} [ \cos(px-ry) + \cos (px + ry) - \cos (rx - py) - \cos (rx + py)] \nonumber \\
&\cdot & \!\!\!\!\! [( \cos(sx-uy) + \cos (sx + uy) - \cos (ux - sy) - \cos (ux -sy)] \nonumber \\
&=& \!\! \![- \sin \frac{1}{2}(px-ry+ rx-py)\sin \frac{1}{2}(px-ry-(rx-py)) \nonumber \\
&-& \!\!\!\!  \sin \frac{1}{2}(px+ry+ rx+py)\sin \frac{1}{2}(px+ry-(rx+py))] \nonumber \\
&\cdot &\!\!\!\!\! [ - \sin \frac{1}{2}(sx-uy+ux-sy)\sin \frac{1}{2}(sx-uy-(ux-sy))\nonumber \\
&-&\!\!\!\! \sin \frac{1}{2} (sx + uy + ux +sy) \sin \frac{1}{2} (sx +uy - (ux+sy))]\nonumber \\
&=&  \!\!\![\sin nv \sin mw - \sin nw \sin mv][ \sin kv \sin lw  - \sin kw \sin lv],
\end{eqnarray}
where
\begin{equation}
   n= p+r, \,\, m = p-r, \,\, k = s+u, \,\, l=s-u, \,\, w = \frac{x+y}{2} \,\,\,\, \text{and}\,\,\,\,   v = \frac{x-y}{2}.
\end{equation}
\end{widetext}
Since both nominator and denominator in (C30) are trigonometric functions we can exploit their periodicity to enlarge the domain of integration and make a shift of variables to $w$ and $v$. An integration over the original domain is namely half the value of an integration over the enlarged domain in fig. 8, thus
\begin{gather}
\int^\pi_{-\pi} \int^\pi_{-\pi} dx dy = \frac{1}{2} \int^{2\pi}_{- 2\pi} dw' \int^{v_2'(w')}_{v_1'(w')} dv'\nonumber \\ =  \frac{1}{2}  \frac{1}{2}  \int^{2\pi}_{- 2\pi} dw'  \int^{2\pi}_{- 2\pi} dv' = \int^\pi_{-\pi} dw \int^\pi_{-\pi} dv,  \nonumber
\end{gather}
\vspace*{-3mm}
\begin{equation}
   \quad \text{where} \quad w' = x+y \quad \text{and} \quad v' =x+y.
\end{equation}
The factor $ \frac{1}{2}$   in the second expression is just from the change of coordinates  $dw'du' = 2dxdy$ and the domain of integration is still not enlarged, but limited by piecewise linear functions $v'_1$  and  $v'_2$ . In the third expression then we double the area of integration to lift the coupling between $w'$  and $v'$. In the last expression we just rescale our variables to suit our needs in (C32). 

With the coordinate transformations in (C34) we can use (C16) to get the final result	
\begin{widetext}	
\begin{equation}
   3 < f_{pqr} \! \mid \! \frac{1}{\sin^2 \frac{1}{2} (x-y)} \! \mid f_{stu} > \nonumber 
\end{equation} 
 \vspace*{-2mm}
\begin{equation}
  = 3 \delta_{qt} \cdot \pi  \int^\pi_{-\pi} dw  \int^\pi_{-\pi} dv \,
     \frac{[\sin nv \sin mw - \sin nw \sin mv] [\sin kv \sin lw - \sin kw \sin lv] }{\sin^2 v} \nonumber
\end{equation} 
\vspace*{-2mm}
\begin{equation}
= 3 \delta_{qt} \cdot \pi [ \delta_{ml} \cdot \pi ( \mid n+k\mid - \mid n-k \mid ) \cdot \pi - 
  \delta_{mk} \cdot \pi ( \mid n+l \mid - \mid n-l \mid ) \cdot \pi \nonumber
\end{equation} 
\vspace*{-3mm}
\begin{equation}
   \hspace*{1cm}  - \delta_{nl} \cdot \pi ( \mid m+k\mid - \mid m-k \mid ) \cdot \pi + 
  \delta_{nk} \cdot \pi ( \mid m+l \mid - \mid m-l \mid ) \cdot \pi ],
\end{equation} \end{widetext}
which is a specific example of the general result (C13).

\section{Toroidal parton distributions}

Inspired by Bettini's elegant treatment of parton scattering [57] we generate distribution functions via our exterior derivative (9). In short the derivation runs like this (with $\hbar = c=1$): Imagine a proton at rest with four-momentum 
$P = ({\text{\bf{0}}}, E_0)$. We boost it to energy $E$ by impacting upon it a massless four-momentum $q = ({\text{\bf{q}}}, E - E_0) $ which we assume to hit a parton $xP$. After impact the parton carries a mass $xE$. Thus
\begin{equation}
    (xP_{\mu} + q_{\mu})\cdot(xP^{\mu} + q^{\mu}) = x^2E^2,
\end{equation}
from which we get the parton momentum fraction
\begin{equation}
   x = \frac{2E_0}{E + E_0}.
\end{equation}
Now introduce a boost parameter
\begin{equation}
    \xi\equiv \frac{E - E_0}{E} = \frac{2 - 2x}{2-x},
\end{equation}
which we shall use to track out an orbit on the u(3) torus ($\xi=1$ corresponds to $x=0$   and vice versa). With the toroidal generator $T$ as introtangling momentum ope\-rator we namely have the qualitative correspondence $q_{\mu} \sim E - E_0 \sim (1-x)E \sim(1-x)T$. That is, we will project on $\xi T \sim(1-x)T$ in order to probe on $xP_{\mu}$. With a probability amplitude interpretation of $\Phi$  we then have
\begin{equation}
  f_T(x)\cdot dx = ( dR_{u = \exp(\theta i T)})^2 \cdot d \theta, \quad \text{where}\quad  \theta= \pi \xi,
\end{equation}
$ f_T(x)$  is the sought for distribution function and $R(\mbox{\boldmath$\theta$})$ is the proton analogue of (44) with the period doublings in (41) mimicked by making $q$ half integer, see (D11) below. The distributions in fig. 7 are traced out with the generators
\begin{equation}
   T_u = \begin{Bmatrix} 2/3 & 0 & 0 \\ 0 & 0 &0 \\0& 0& -1 \end{Bmatrix} \hspace{2mm} \text{and} \hspace{2mm}
     T_d = \begin{Bmatrix} -1/3 & 0 & 0 \\ 0 & 0 &0 \\0& 0& -1 \end{Bmatrix}\!.
\end{equation}
With (D3) in (D4) we have in general
\begin{equation}
  x \cdot f_T(x) = x \cdot dR^2_{\exp(\frac{2-2x}{2-x}i \pi T)}\cdot \frac{\pi \cdot 2}{(2-x)^2}.
\end{equation}		

The peak position of the distribution functions gene\-rated by (D6) depends on the 'direction' of $T$ relative to $(T_1,T_2,T_3)$. The choices (D5) are 'skew' fractionations of the charge operator $Q_2$ in (B11) which in a matrix representation reads
\begin{equation}
  Q_2/ \hbar = \begin{Bmatrix} 1&0&0  \\ 0&0&0 \\ 0&0&-1 \end{Bmatrix}\!.
\end{equation}			

Since the tracking in (D6) is done from laboratory space we project on a fixed basis like in (A8) when we calculate the exterior derivative $dR$. Any toroidal tracking generator $T$ is a tangent vector and can be expressed in a fixed basis of the tangent space of the torus (the Cartan subalgebra). We write
\begin{equation}
   T = a_1T_1 +  a_2T_2 +  a_3T_3, 
\end{equation}
where $a_j$  are the coefficients of expansion and the $T_j$'s  are the toroidal generators also used in (4). In quark language in the end we want to project $dR$ on all three colour degrees of freedom - cf. (A4) - and sum the contributions to get the total amplitude for the specific flavour amplitude along the specific tracing generator. In general we have [14, 58, 59, 60, 61] (see appendix E)
\begin{gather}
  \sum^3_{j=1} dR_u(\partial_j)= \sum^3_{j=1} dR_u(uiT_j)  \nonumber \\
  =\sum^3_{j=1} dR_u(d \exp  {\text{\Large{$\mid$}}}_{(\theta_1, \theta_2, \theta_3)} 
  ( \frac{\partial}{\partial  \theta_j}))   \nonumber\\ 
   =
    \sum^3_{j=1} dR^{*}_{(\theta_1, \theta_2, \theta_3)} ( \frac{\partial}{\partial \theta_j}) =
   \sum^3_{j=1} \frac{\partial R^{*}}{\partial \theta_j} {\text{\Large{$\mid$}}}_{(\theta_1, \theta_2, \theta_3)},
\end{gather}
where $ u= \exp(\theta_1iT_1 + \theta_2 i T_2 + \theta_3iT_3)$ and $R^{*}$ is the pull-back of $R$ to parameter space from the manifold. We shall immediately omit the asterix on $R^{*}$ again and we have finally
\begin{gather}
   dR_{u = \exp(\theta i T)}( \partial_1 + \partial_2 + \partial_3)  \nonumber \\
 =\left(  \frac{\partial R}{\partial \theta_1}+ \frac{\partial R}{\partial \theta_2}+ \frac{\partial R}{\partial                    \theta_3} \right){\text{\Large{$\mid$}}}_{(\theta_1, \theta_2, \theta_3)
  =(\theta \cdot a_1, \theta \cdot a_2, \theta \cdot a_3)}     \nonumber \\
 \equiv D(\theta \cdot a_1, \theta \cdot a_2, \theta \cdot a_3) \equiv D_T(\theta),
\end{gather}
where we have introduced a shorthand notation for the directional derivative.

In our specific case in fig. 7 we use a first order term (C2) from (44) with  $(p,q,r)=(0, \frac{1}{2},1)$
\begin{equation}
  b_{0, \frac{1}{2},1}  (\theta_1, \theta_2, \theta_3)  = \frac{1}{N} \begin{vmatrix}  1&1 & 1\\
\sin \frac{1}{2} \theta_1 & \sin \frac{1}{2} \theta_2 & \sin \frac{1}{2} \theta_3 \\
 \cos \theta_1 & \cos \theta_2 & \cos \theta_3\end{vmatrix}\!,
\end{equation}
where $N$ is a normalization constant. For the particular case of (D11) we get a specific expression for the function $D$ introduced in (D10)
\begin{align}
ND(\theta_1, \theta_2, \theta_3) = \hspace{6cm} \nonumber \\
- \frac{1}{2} \cos \frac{\theta_1}{2}\cdot (\cos \theta_3 - \cos \theta_2) - \sin \theta_1 \cdot (\sin \frac{\theta_3}{2}-\sin \frac{\theta_2}{2} ) \nonumber \\
+  \frac{1}{2} \cos \frac{\theta_2}{2}\cdot (\cos \theta_3 - \cos \theta_1)+  \sin \theta_2 \cdot (\sin \frac{\theta_3}{2}-\sin \frac{\theta_1}{2} )\nonumber \\
-  \frac{1}{2} \cos \frac{\theta_3}{2}\cdot (\cos \theta_2 - \cos \theta_1) - \sin \theta_3 \cdot (\sin \frac{\theta_2}{2}-\sin \frac{\theta_1}{2} ).   
\end{align}
And tracking with $T_u$ from (D5) we get the final result for our distribution function
\begin{align}
   xf_u(x) = \hspace{6.5cm} \nonumber \\
x \left[D(\theta \cdot \frac{2}{3}, \theta \cdot 0 , \theta\cdot(-1))\right]^2 \cdot {\text{\Large{$\mid$}}} \frac{d \theta}{dx} {\text{\Large{$\mid$}}} , \,\, \text{where} \,\, \theta \equiv \pi \xi.\end{align}
Expressed in the parton fraction $x$ (D13) reads
\begin{widetext}
\begin{equation}
  xf_u(x)=  
x\left[D\left( \pi \frac{2 - 2x}{2-x} \cdot \frac{2}{3}, \pi  \frac{2 - 2x}{2-x} \cdot 0, 
   \pi \frac{2 - 2x}{2-x} \cdot (-1) \right) \right]^2 \cdot \frac{\pi \cdot 2}{(3-x)^2}.
\end{equation}
\end{widetext}
The result (D14) is shown in fig. 7 together with the result of tracking with $T_d$. Note that $x \cdot f_{Tu}(x)\mid_{max} \approx 0.63$  and $x \cdot f_{Td}(x)\mid_{max} \approx 0.32$   in rather fine agreement with u$_v$ and d$_v$ in the distribution functions shown in fig. 7, and which are experimentally extracted via NLO-QCD parametrizations. Note also that
\begin{gather}
   \int^1_0xf_{Tu}(x) dx \, {\text{\Large{/}}} \int^1_0xf_{Td}(x) dx \nonumber \\ = 0.2722 {\text{\Large{/}}} 0.1437  = 
 1.89 \approx 2 ,\end{gather}
which for a first order approximation (D11) is in rather close agreement with the relative constituent quark flavour model content of the proton according to the standard model, namely 2u:1d. Actually the ratio between the distributions themselves yields

\begin{equation}
    \int^1_0f_{Tu}(x) dx \, {\text{\Large{/}}} \int^1_0f_{Td}(x) dx = 1.33\ldots {\text{\Large{/}}} 0.66 = 1.996 \ldots .
\end{equation}

We should also note that when comparing with established distribution functions one should be aware that these are slightly scale dependent as seen when comparing those for 10 GeV$^2$ with those for 10,000 GeV$^2$ from the particle data group [62]. One should expect closer correspondence the lower the energy scale. Actually we would like to see data extracted at $\approx$~0~GeV$^2$, which would represent the purely allospatially derived distribution functions. Luckily though, it seems that 10 GeV$^2$ is small enough to make the bridge. We could call it the bridge of soft 'deep' inelastic scattering. The softer the momentum transfer gets the more terms one needs in the QCD-expansions through which the data are currently analyzed. The distributions in the insert in fig. 7 are parametrizations from next to leading order, NLO calculations. What we would need in order to reach allospatial dynamics would be N$^{\infty}$LO-QCD  at $Q^2$ =~0~GeV$^2$. This would be the confinement domain which we have attacked in the present work 'from within', but which is the domain where pertubative QCD is expected to break down, i.e. where N$^{n}$LO-QCD is not expected to converge for $n \rightarrow \infty$.

\section{Vector fields, derivations and forms on smooth manifolds}

To understand the use in (A6) of left invariance of the coordinate fields and linearity of the coordinate forms actually requires a careful application of the different expressions for the directional derivative. And to arrive from the first expression in (D9) to the last expression of the same equation specifically requires a consistent definition of differentiation on smooth mani\-folds. In the present appendix we shall outline the basic concepts. The appendix is based almost exclusively on [58]. We have only added two corollaries of particular relevance in the present context. Here it comes:

Let $Z$ be the generator of a directional derivative of $\Psi$ at a certain point $u$ in our manifold u(3). Repeating eq. 8 of [14] we then have the following equivalent expressions for the directional derivative
\begin{equation}
   Z_u \left[\Psi\right] \equiv  Z \left[\Psi\right] \!(u) = (d \Psi)_u (Z) = \frac{d}{dt} \Psi(u \exp(tZ)) \!\mid_{t=0} \! .
\end{equation}
The identity between the second and the third expression in (E1) is just a question of notation. 

Referring to fig. 9 we namely have for the directional derivative of real functions $f$ on a smooth manifold $M$ with tangent space $TM$ the following
\begin{gather}
	\text{\it{Definition 1:}} \nonumber \\ 
	 \text{Let}  \,\, f: M \rightarrow R \,\, {\text{be a smooth function and}} \,\,X_p \in      TM_p. \nonumber \\ {\text{The number}}\,\, df_p(X_p)\,\, {\text{is called}} 
	{\text{ the (directional) derivative}}\nonumber \\{\text{of}}\, f \,{\text{in the direction}}\,\, X_p \,\, {\text{and is also denoted}} \,\,  X_p(f).
\end{gather}										
The directional derivative $X_p(f)$ fulfils the following identities
\begin{gather}
  X_p(f+g)=X_p(f)+X_p(g)  \qquad f,g \in C^{\infty}(M) \\
  X(\lambda \cdot f) = \lambda \cdot X_p(f)\qquad  \lambda \in R\\
  X_p(f\cdot g) = X_p(f) \cdot g(p) +f(p)\cdot X_p(g).
\end{gather}
 where $f+g$   and $f\cdot g$  are the functions given by addition and multiplication of functional values. We see that the directional derivative applies and acts algebraically in analogy to the usual act of differentiation.  

We now state the following
\begin{gather}
	\text{\it{Definition 2:}}  \nonumber \\ 
	 {\text{A map}}  \,\, D_p: C^{\infty}(M) \rightarrow R \,\, {\text{which fulfils (E3)-(E5)}} \nonumber \\
 {\text{is called a derivation at the point}}\,\, p.
\end{gather}
We see that any tangent vector gives rise to a derivation. It is remarkable and important for our purpose, that the opposite is also true:
\begin{gather}
	\text{\it{Theorem 1:}}  \nonumber \\ 
	{\text{Any derivation}}\,\, D_p \,\,{\text{at the point $p$}} \nonumber \\ {\text{ is a directional derivative.}} 
\end{gather}
This theorem is used in order to write the identity between the third and the fourth (last) expression in (E1). We shall leave out the proof of (E7) which needs some lemmas that would lead us astray in the present context. The conclusion is that there exists a 1-1 correspondence between tangent vectors $X_p \in TM_p$   and derivations $D_p:C^{\infty}(M) \rightarrow R$ at the point $p$: The derivation corresponding to $X_p$  is
\begin{equation}
   X_p(f) = df_p(X_p) \quad {\text{or}}
\end{equation}
\begin{equation}
   X_p(f) =  \sum_i a_i \frac{\partial f}{\partial x_i} \! \mid_p,
\end{equation}
where $X_p= \sum_i a_i \frac{\partial}{\partial x_i}\! \!\mid_p$ and $ \frac{\partial f}{\partial x_i} \!\!\mid_p = \frac{(f \circ x^{-1})}{\partial x_i}\!\!\mid_{x(p)}$  for an atlas  $(U,x)$, that covers the manifold $M$. 

\begin{figure} [h]
\begin{center}
\includegraphics[width=0.4\textwidth]{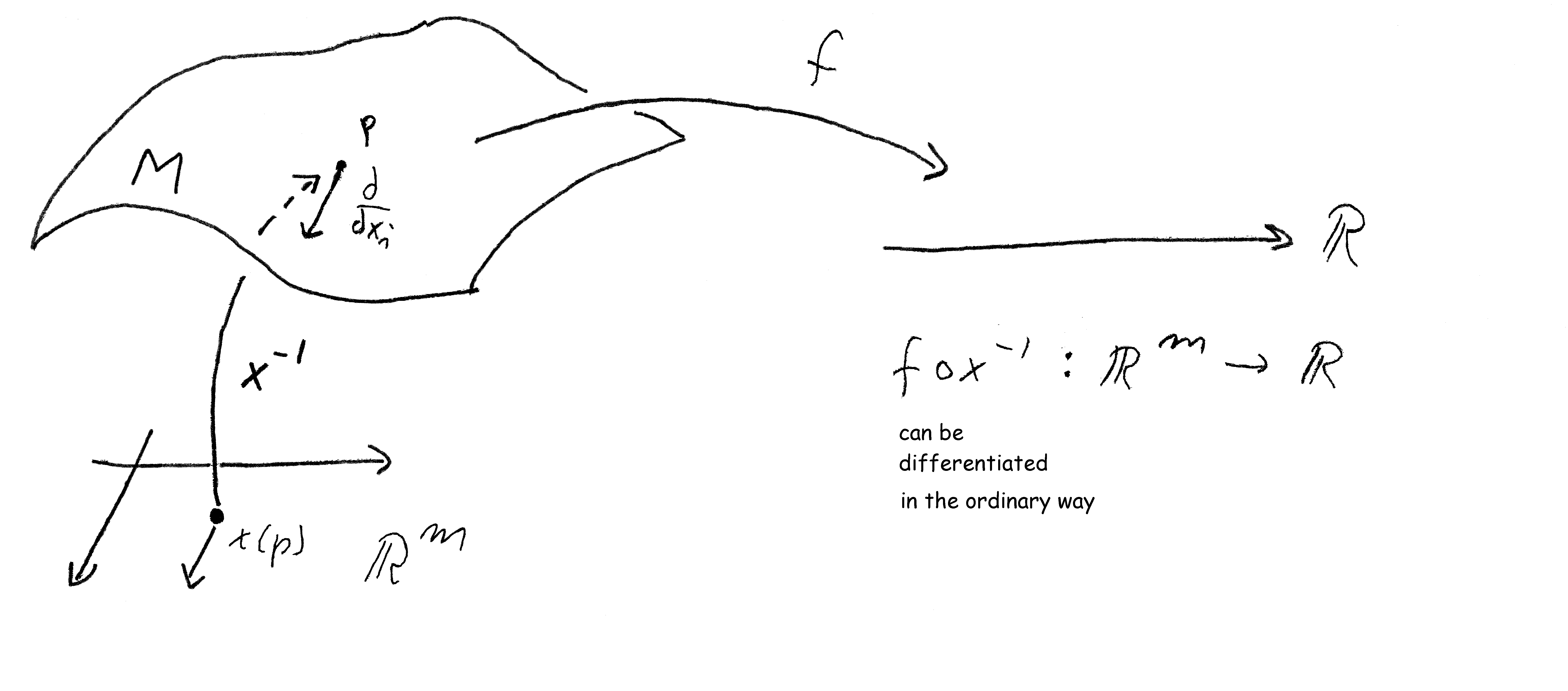}
\caption{Derivation of a real function $f:M\rightarrow R$ at point $p$ in the mainfold $M$ is defined by using a local smooth map $x:M\rightarrow R^m$ to pull back the problem to an ordinary derivation on $R^m$ by using the pullback function $f \circ x^{-1}:R^m\rightarrow R$. One can differentiate $f \circ x^{-1}$ in the ordinary way. This idea is readily generalized to a complex valued function and in our case the manifold $M$ could be u(3) and the complex valued function $f$ could be either the wavefunction $\Psi$ or its measure scaled partner $\Phi$.}
\label{figE1}
\end{center}
\end{figure}

The idea is illustrated in fig. 9 locally for a particular local map in the book of maps, the atlas, which consist of enough maps to parametrize all of $M$ via local maps that are joined smoothly in their overlapping regions. The manifold $M$ is depictured in fig. 9 as a surface in an embedding space of specific dimension. So now in short we have a global definition of derivation and can faithfully derive locally by pulling back the derivation on the manifold to ordinary derivations in a parameter space $R^m$  of the same dimension $m$ as the manifold $M$. Note that the presence of the spikes in the parametric potential (17) shown in fig. 2 does {\it{not}} signify that the geodetic distance (3) used for our potential (2) in our defining Hamiltonian (3) would not be smooth. Quite the contrary: The geodetic distance, and thus its square, {\it{are}} smooth functions on all of u(3). The spikes in fig. 2 only indicate that one cannot define a single map to cover u(3) globally ("You cannot peel an orange without breaking the skin"). This is also why we have to limit the expression (2) to the interval $- \pi \le \theta_j \le \pi$  for our parametrization which we need to make our Hamiltonian operational for actual calculations in (16). But luckily we {\it{can}} use a common parametrization of the eigenvalues of $u \in u$(3). 

To proceed further we need the following
\begin{gather}
	\text{\it{Statement 1:}}  \nonumber \\ 
	{\text{The smooth vector fields}}\,\,  \Gamma (M) \,\,{\text{on}}\,\, M \,\, {\text{constitute}} \nonumber \\ {\text{an (infinite dimensional) real
    vector space.}} 
\end{gather}
My personal notes in the lecture notes [63] explains the infinity with these comments: "There is an 
infinity of atlases and infinitely many coefficients to $\frac{\partial}{\partial x_i}\!\! \mid_p $. At any point  $ p \in M $ there is an $m$-dimensional infinity of tangent vectors, (and) these can be combined in infinitely many ways at different $p$'s." To understand the statement (E10) we need some more math. First let  $M^m \subset R^k$  be a smooth manifold. We shall then start out with the following [64]
\begin{gather}
	\text{\it{Definition 3:}}  \nonumber \\ 
	{\text{A function}}  \,\, X:M^m \rightarrow R^k \,\, {\text{which fulfils}}  \,\,
       X(p) \in TM_p \nonumber \\{\text{for all}} \,\, p \in M\,\, {\text{is called a vector field on}}\,\, M. \,\, {\text{If}} \,\, M \,\, {\text{is}} \nonumber \\ {\text{continuous (smooth), the vector field is said to be}} \nonumber \\
 {\text{continuous (smooth).}}
\end{gather}

Please allow me to cite my teacher's comment on the above definition: "Properties of vector fields and their dynamical systems on manifolds are studied locally by translating the previously treated euclidean case by the help of atlases." This is a key point in our application in (D9) of the beautiful mathematical machinery exposed in this appendix. Let us now continue with his explanation following my citation: Let $(U,x)$  be a chart on $M^m$. Corresponding to this we have an {\it{induced base}} for any point $p \in U$:
\begin{equation}
    \frac{\partial}{\partial x_i}\!\! \mid_p = d(x^{-1})_{x(p)} (e_i),
\end{equation}
where $d(x^{-1})_{x(p)} : R^m \rightarrow TM^m_p \subset R^k$  
and where $e_i$  is the $i$'th standard unit vector in  $R^m$. Since $e_1, \ldots,e_m$  is a base for   $R^m$
\begin{equation}
  \frac{\partial}{\partial x_1}\!\!\mid_p , \ldots , \frac{\partial}{\partial x_m}\!\!\mid_p
\end{equation}
 becomes a basis for $TM_p^m$. I made a remark in his notes that "$d(x^{-1})$  is injective since $x$ is a diffeomorphism" (per definition of the smoothness of $M$, OT 2012). Perhaps we should have mentioned earlier, that $U \subset M$ is an open neighbourhood of $p \in U$, which is where we have defined (directional) derivatives above (E2).

Going back through the necessary mathematical concepts for our purpose, we give for the sake of completeness: 
\begin{gather}
	{\it{Definition \, 4:}}  \nonumber \\ 
 {\text{Let}}  \,\, p \in M^m \subset R^k \,\, {\text{be a point on a smooth}}  \,\,m
       {\text{-dimensional}}\nonumber \\ {\text{manifold and}} \,\, (U,x)\,\, {\text{a }} 
 {\text{chart around}}\,\, p \,\, {\text{with}} \,\, x(p)=\alpha. \nonumber \\ {\text{The tangent space}}\,\,  TM_p \,\,{\text{for}} \,\, M^m \,\, {\text{ in the point $p$ is the}}\nonumber \\
 {\text{image of the linear map}}\nonumber
\end{gather}
\vspace{-8mm}               
\begin{equation}
  d(x^{-1})_\alpha :R^m \rightarrow R^k.
\end{equation}
\begin{gather}
	{\text{\it{Definition 5:}}}  \nonumber \\ 
	 {\text{Let}}\,\, M^m \in R^k \,\, {\text{and}}  \,\,N^n \in R^l\,\,
       {\text{be smooth manifolds and}} \nonumber \\
   f:M^m \rightarrow N^n  {\text{ a smooth function. By the differential of}} \nonumber \\ f {\text{ in the point}} \,\, p\,\, {\text{is understood the linear map}}\nonumber \\
d f_p : TM_p \rightarrow TN_{f(p)}\\
 {\text{defined in the following way: Let}}\,\, V\,\, {\text{be a neighbourhood }} \nonumber \\ {\text{of}}
  \,\, p \,\, {\text{in}} \,\, R^k \,\, {\text{and}}\,\, \overline{f} : V \rightarrow R^l 
{\text{ a smooth continuation of $f$}}\nonumber \\ ({\overline{f}} \mid V \cap M=f) \,\, {\text{(or in words:}}
  \,\, \overline{f} \,\, {\text{'s restriction to }} M \nonumber \\
 {\text{is}} \,\, f, \,\, {\text{OT 2012). Then for}} \,\, \xi \in T M_p \,\, {\text{the diffential is}}
  \nonumber \\
 d f_p(\xi) = d \overline{f}_p(\xi).
 \end{gather}
 
We are now ready to explain statement 1 in (E10), which we shall need for essential algebraic manipulations below. Statement 1 might be obvious in connection with the lecture notes, but here we need to strengthen the statement by formulating the following
\begin{gather}
	{\text{\it{Corollary 1:}}}   \nonumber \\ 
	{\text{Let}}\,\,df:TM^m \rightarrow TR=R \,\, {\text{be the diffential (the exterior}} \nonumber \\ {\text{derivative) of a function}}   
  f : M^m \rightarrow R. \,\,\, {\text{Further let}} \nonumber \\
X _p(f) = \sum_i a_i \frac{\partial f}{\partial x_i} \!\!\mid_p \quad  {\text{and}} \\
     Y _p(f) = \sum_i b_i \frac{\partial f}{\partial x_i} \!\!\mid_p  \\
 {\text{be (directional) derivatives of}} \,\, f  {\text{ in the directions}} \,\, X_p\, \nonumber \\ {\text{  respectively}}\,\, Y_p.\,\, {\text{Then}} \nonumber \\
Z_p (f) \equiv \sum_i c_i \frac{\partial f}{\partial x_i}\!\! \mid_p, \quad \quad {\text{where}} \,\, c_i \equiv a_i + b_i, \\
{\text{will be a directional derivative in the direction}} \,\, Z_p. \nonumber \\ {\text{ In other words}} \nonumber \\
( X_p+Y_p)(f)=X_p(f)+Y_p(f).  
 \end{gather}

{\it{Proof of corollary 1:}}\\
It should be clear that $TM^m_p$  is a linear vector space and thus it is clear that $Z_p \in TM_p$. We also remarked in the text following (E6), that $Z_p$  will be a derivation. This is obvious from definition 1 in (E2) when comparing with definition 2 in (E6).  Now from theorem 1 in (E7) we then have that  $Z_p$ will also be a (directional) derivative. Q.E.D.
It shall be essential to us that $Z_p$  in (E19) is expanded on the {\it{same basis}} as that of $X_p$ and $Y_p$ . This is possible because one can use the same smooth continuation $\overline{f}$   for all the three derivatives  $X_p, \, Y_p$  and $Z_p$. We need a second
\begin{gather}
	{\text{\it{Corollary 2:}}}   \nonumber \\ 
	{\text{Let}}\,\,f: M^m \rightarrow R \quad {\text{then}}  \nonumber \\ 
    df(X+Y)=df(X) + df(Y), \\  {\text{where}} \quad X,Y \in TM^m    \nonumber \\
   {\text{and}}\quad d f(\lambda X) = \lambda \cdot df(X),\\  {\text{where}} \quad \lambda \in R. \nonumber
 \end{gather}
Here we have suppressed the index $p$, but (E21) and (E22) should be understood pointwise. It should be obvious that (E21) in corollary 2 is just a formal way of expressing the statement in definition 5, that $df_p$  is a linear map, and that (E22) in corollary 2 is just another way of stating one of the characteristics of derivatives, namely (E4). However, we have not proven the linearity of forms, which has been the formulation (exterior derivative) primarily used in the main text of the present work. Thus we wanted to state this small
\begin{gather}
{\text{Proof of corollary 2:}}\nonumber \\
   df_p(X+Y) = (X_p+Y_p)(f) =  \nonumber \\
X_p(f) + Y_p(f) = df_p(X) + df_p(Y). 
\end{gather}
In the second expression we used the equivalent notation for directional derivatives introduced in definition 1 in (E2), in the third expression we used (E20) and in the last expression we returned to the language of forms via (E2). To prove (E22) we write
\begin{equation}
   df_p(\lambda X) = \lambda X_p(f) = \lambda d f_p(X).
\end{equation}
In the second expression we used again definition 1 in (E2). The last expression finishes the proof. Q.E.D

Of course the "Proof of corollary 2" is not really a proof, there was nothing new to prove. The 'proof' is only meant to show how the algebra works when one switches between the different notations $X_p(f)$   and $df_p(X)$  for the directional derivative.

Corollaries 1 and 2 are readily generalized to $f:M^m \rightarrow C$  and $\lambda \in C$. With these two corollaries we now have our mathematical machinery in shape to proceed from the first to the last expression in (D10) in order to calculate (D12). 

The linearity of forms expressed in corollary 2 by the eqs. (E21) and (E22) together with the generalization (E2) of derivation in definition 1 and finally the existence of induced bases like in (E12) pull back the problem to $R^m$  so that the partial derivatives pull back to give results like (D12). We have thus presented the derivation on the smooth manifold u(3) in an operational form which makes possible actual calculations.

\newpage

\begin{widetext}
\section*{Tables}
\begin{table} [ht!]
\begin{center}
\caption{Comparison of the 10 lowest lying eigenvalues E of (1)/(22) for isospin $\frac{1}{2}$ and hypercharge 1. The eigenvalues are given without and with the curvature and centrifugal potentials from a Rayleigh-Ritz calculation with 1800 base functions of the type (44)  corresponding to $p = 0, 1, 2,..., 14;\ q = 1,2,...,15;\ r = p+1,...,15$ . Minimum values have been used for $\mathbf{K}^2$ and $\mathbf{M}^2$, namely $K(K+1)=\frac{1}{2}(\frac{1}{2}+1)$  and $M^2=\frac{13}{4}$ , following the interpretation in (B14). The corresponding shifts in energy between calculations without and with the curvature and centrifugal potentials lie within 50 MeV as claimed in the main text. The somewhat curious order in the list of the alleged candidates follows from choosing them according to the eigenvalue of the charged partners from approximate solutions in sect. 4. The value 1723 MeV in brackets in the second column is the eigenvalue of such a charged partner. \vspace{3mm}}
\label{tab:lowest10}
\begin{tabular}{c c c c c c} \hline\hline\\
Eigenvalues & \textit{E} & Eigenvalues & \textit{E} & Candidate & Relative\\
no curvature & & curvature & & & shift from\\ no centrifugal (31) & & centrifugal (22) & & & (31) to (22)\\
 & MeV & & MeV & &
\\ \hline\\
4.47&939.6 (fit)&4.38&939.6 (fit)&n, p&-2.0 \\
6.22&1308&6.10&1309&N(1440)&-1.9 \\
6.57&1381&6.53&1401&N(1535)&-0.6 \\
8.19&1722&8.12&1742&N(1675)&-0.9 \\
8.32&1749&8.42&1806&N(1720)&1.2 \\
8.39&1764&8.53&1830&N(1680)&1.7 \\
9.17&1928 (1723)&9.35&2006&N(1650)&2.0 \\
10.29&2163&10.37&2225&N(2250)&0.8 \\
10.36&2178&10.54&2261&N(2190)&1.8 \\
10.48&2203&10.60&2274&N(2220)&1.1 \\ \hline\hline
\end{tabular}
\end{center}
\end{table}

\begin{table} [ht!]
\begin{center}
\caption{Eigenvalues of the parametric group space chopped harmonic oscillator Schrödinger equation (32) calculated with 1500 collocation points. The same results are obtained by 1) iterative integration, 2) McLaurin series [65] and 3) collocation, see table 6. The mutual discrepancies are at the level of 10$^{-8}$. This fine agreement among the different methods lends support to the Rayleigh-Ritz method also for solving the full eq. (22). Note that the lowest eigenvalues as expected are close to those of the ordinary harmonic oscillator since the lowest states live in the neighbourhood of origo and thus do not 'feel' that the potential is chopped into periodicity. Moving up to higher levels the eigenvalues differ more and more from those of the harmonic oscillator as indicated in fig. 5. \vspace{3mm}}
\label{tab:paraeig}
\begin{tabular}{ c c c c } \hline\hline\\
Level&Eigenvalue&Diminished&Augmented \\ \hline \\
1&0.4998047079793375&&0.5001727903915900 \\
2&1.502988968183189&1.496433950157817& \\ 
3&2.471378779213570&&2.522629649224744 \\
4&3.600509000413400&3.377236031951678& \\
5&4.218515963091988&&4.803947526779894 \\
6&6.197629004032325&5.160535373287425& \\
7&6.383117406428158&&7.820486992163699 \\
8&9.688466291114409&7.922699153838795& \\ 
9&9.751335596178837&&11.80644675634627 \\
10&14.17552754458349&11.84897047349423& \\
11&14.20637080732072&&16.79575229446833 \\ \hline\hline
\end{tabular}
\end{center}
\end{table}

\begin{table} [ht!]
\begin{center}
\caption{Scarce singlet states. Eigenvalues based on Slater determinants of three cosines up to order 20 analogous to (44). The first column shows eigenvalues of the approximate eq. (31) and the third column shows eigenvalues of the exact eq. (22). A singlet 579-like resonance is predicted at 4499 MeV in the free charm system $\Sigma_c^+(2455)D^-$  slightly above its threshold at 4324 MeV. It should be visible in neutron diffraction dissociation experiments like those in ref. [37]. The rest masses are predicted with the common fit of table 1 where 939.6 MeV corresponds to the ground state N. \vspace{3mm}}
\label{tab:singlets}
\begin{tabular}{c c c c c} \hline\hline \\
Singlet & Toroidal & Singlet & Toroidal & \phantom{xx}Rest mass\phantom{xx}\\
approximate (31) & label & exact (22) & partner & MeV/c$^2$\\ \hline \\
7.18957249&1 3 5&7.12174265&1 3 5 &1526 \\ 
9.35680923&1 3 7&9.57104964&1 3 7&2051 \\ 
11.11924735&1 5 7&11.29403818&1 5 7&2420 \\ 
12.71754526&1 3 9&13.25048053&1 3 9&2839 \\ 
13.09274266&3 5 7&13.28113847&&2846 \\ 
14.44940740&1 5 9&14.96408154&1 5 9&3206 \\ 
16.40861790&3 5 9&16.92132768&&3626 \\ 
16.66054860&1 7 9&17.30060283&1 7 9&3707 \\
17.17694213&1 3 11&18.00904018&1 3 11&3859 \\
18.63196975&3 7 9&19.25767851&&4126 \\
18.92139903&1 5 11&19.73271915&1 5 11&4228 \\
20.37744401&5 7 9&20.99400214&5 7 9&4499 \\
20.89101474&3 5 11&21.71097283&3 5 11&4652 \\
21.07660092&1 7 11&22.04093052&1 7 11&4723 \\ \hline\hline
\end{tabular}
\end{center}
\end{table}

\begin{table} [ht!]
\begin{center}
\caption{Consistency of results from expanding on Slater determinants of 1-dimensional functions like (33) and (44) to solve the full eq. (22). In the first column are the lowest lying N-states constructed from two cosines and one sine as in (44). In the second and third columns are the lowest lying states constructed from parametric eigenfunctions as in (33). Note that the states based on parametric eigenfunctions cover the N-states fully and further include also states that are identified as either singlet or $\Delta$-states. The parametric base eigenvalues use the eleven lowest lying solutions to (32) with respectively 47$^3$ and 29$^3$ base points for each function. The rest masses in the last column are predicted from a fit to the ground state of the parametric base where 939.6 MeV corresponds to the ground state eigenvalue 4.364. \vspace{3mm}}
\label{tab:consistency}
\begin{tabular}{c c c c c c} \hline\hline \\
Trigonometric & Parametric base & & & Toroidal & \phantom{xx}Rest mass\phantom{xx}\\
base & eigenvalues & & & labels & \\ 
N-states & 47$^3$ & 29$^3$ & & & Mev/c$^2$ \\ \hline \\
4.385&4.364&4.345&&& 939.6 (fit) \\ 
&5.568&5.546& $\Delta$-state &124&1199 \\ 
6.103&6.087&6.067&&& \\
6.537&6.517&6.493&&& \\
&7.054&7.033&Singlet&135&1519 \\
&7.538&7.508& $\Delta$-state&234&1623 \\ 
8.119&8.087&8.060&&& \\ 
8.424&8.368&8.333&&& \\ 
8.533&8.513&8.476&&& \\ 
9.356&9.326&9.290&&&2008 \\ \hline\hline
\end{tabular}
\end{center}
\end{table}

\begin{table} [ht!]
\begin{center}
\caption{Spin and parity conjecture on allospatial states compared with the series of all the observed 4-star N-resonances. The assignments are only tentative. An alternative assignment around the singlet state 137 is shown in brackets.\vspace{3mm}}
\label{tab:spinparity}
\begin{tabular}{c c c c c} \hline\hline\\
$(-1)^{n-l};\ l,m,n$ & $J^P$ & $L_{2I,2J}$ & Name & RPP rating\\ \hline\\
+ 123&1/2+&P11&n, p&**** \\
+ 125&1/2+&P11&N(1440)&**** \\ 
-134&1/2-&S11&N(1535)&**** \\
+(135)&1/2-&S11&N(1650)&**** \\ 
&&&& \\ 
- 235&3/2-&D13&N(1520)&**** \\ 
+145&3/2+&P13&N(1720)&**** \\ 
+ 127&5/2+&F15&N(1680)&**** \\ 
-136&5/2-&D15&N(1675)&**** \\ 
&&&& \\ 
+ (137)&&&N(2040)&- \\ 
&&&& \\ 
+ 345&(9/2-)&(G19)&(N(2250))&(****) \\ 
- 237&7/2-&G17&N(2190)&**** \\ 
+ 147&9/2+&H19&N(2220)&**** \\ 
- 156&9/2-&G19&N(2250)&**** \\ 
...&&&& \\
-257&11/2-&I111&N(2600)&*** \\ 
 \hline\hline
\end{tabular}
\end{center}
\end{table}

\begin{table} [ht!]
\begin{center}
\caption{Comparison of numerical results for the eigenvalue of the ground state. The seperable problem (31) has been solved by four different methods three of which gives a set of eigenvalues for the one-dimensional problem (32) from which the eigenvalues for the three-dimensional problem (31) is constructed. These eigenvalues can be used to check the Rayleigh-Ritz method for solving the three-dimensional problem directly. Mutual discrepancies are due to the finite expansions in the different methods. The fine agreement among the different methods lends support to the Rayleigh-Ritz method also for solving the full eq. (22).\vspace{3mm}}
\label{tab:numericalmethods}
\begin{tabular}{c c c c c c} \hline\hline \\
1D-level & Iterative & MacLaurin & Rayleigh-Ritz & Collocation & Collocation\\
number & integration & series [65] & 1800 & 1500 & 1500\\
& & & base functions & points & points\\ \\
& Comal & - & MathCad & Matlab $m_n$& Matlab $m_p$ \\ \hline\\
1&0.499804708&0.499804704&-&0.499804708&0.500172790 \\ 
2&1.502988981&1.502988968&-&1.502988968&1.496433950 \\
3&2.471378882&2.471378899&-&2.471378779&2.471378779 \\ 
Sum&4.474172571&4.474172571&4.47417271&4.474172455&4.467985519 \\ \hline\hline
\end{tabular}
\end{center}
\end{table}
\end{widetext}

\clearpage

\section*{{References}}
\hspace{-3.5mm}1. E. Klempt and J. M. Richard, {\it{Baryon Spectroscopy}}, Rev.\ Mod.\ Phys.\ {\bf{82(2)}}, 1095, (2010).\newline
2. R.\ A.\ Arndt et.\ al., {\it{Extended partial-wave analysis of $\pi N$scattering data}}, 
Phys. Rev. {\bf{C74}}, 045205 (2006). \newline
3.\ K. Nakamura et.\ al. (Particle Data group) 2010, {\it{Review of Particle Physics}}, J. Phys. G: Nucl. Part. Phys {\bf{37 (7A)}}, 075021 (2010), p.189. \newline
4. K. Nakamura et.\ al., op. cit., pp.202. \newline
5.\ R. J. Holt and C. D. Roberts, {\it{Nucleon and pion distribution functions in the valence region}}, Rev. Mod. Phys {\bf{82(4)}}, 2991-3044, (2010). \newline
6. A. Bazavov et. al., {\it{Nonpertubative QCD simulations with 2+1 flavors of improved staggered quarks}}, Rev.\ Mod.\ Phys.\ {\bf{82(2)}}, 1349-1417 (2010). \newline
7. C. Alexandrou, R. Baron, J. Carbonell, V. Drach, P. Guichon, K. Jansen, T. Korzec and O. Pène, {\it{Low-lying baryon spectrum with two dynamical twisted mass fermions}}, arXiv: 0910.2419v1 [hep-lat] (2009).\newline 
8. S. J. Brodsky, H.-C. Pauli and S. S. Pincky, {\it{Quantum chromodynamics and other field theories on the light cone}}, Phys. Rep. {\bf{301}}, 299-486 (1998). \newline
9. J. Milnor, {\it{Morse Theory}}, Ann. of Math. Stud. {\bf{51}}, 1 (1963).  \newline
10. Hans Plesner Jacobsen, Department of Mathematics, Copenhagen University, Denmark, private communication (approx.\ 1997).\newline
11.\ K. G. Wilson, {\it{Confinement of quarks}}, Phys. Rev. {\bf{D10}}, 2445-2459 (1974). \newline
12.\ N.\ S.\ Manton, {\it{An Alternative Action for Lattice Gauge Theories}}, Phys.\ Lett.\ {\bf{B96}}, 328-330 (1980).\newline
13. Particle Data Group, Eur. Phys. J. {\bf{C3}}, 1 (1998). \newline
14.\ O. L. Trinhammer and G. Olafsson, {\it{The Full Laplace-Beltrami operator on U(N) and SU(N)}}, arXiv: 9901002 [math-ph] (1999). \newline
15. J. S. Dowker, {\it{Quantum Mechanics on Group Space and Huygens' Principle}}, Ann. of Phys. {\bf{62}}, 361-382 (1971). \newline
16. L. I. Schiff, {\it{Quantum Mechanics}}, 3$^{\rm{rd}}$  ed. (McGraw-Hill 1968), p.209. \newline
17. M.\ Gell-Mann, {\it{Symmetries of Baryons and Mesons}}, Phys.\ Rev.\ {\bf{125(3)}}, 1067-1084, (1962). \newline
18. L. Fonda and G.C. Ghirardi, {\it{Symmetry Principles in Quantum Physics}} (Marcel Dekker, New York 1970), p.171.\newline 
19. S. Gasiorowicz, {\it{Elementary Particle Physics}} (Wiley and Sons, New York, 1966) p.261. \newline
20.\ S. Okubo, {\it{Note on Unitary  Symmetry in Strong Interactions}}, Prog. Theor. Phys. {\bf{27(5)}}, 949-966, (1962). \newline
21. Y. Ne'eman, {\it{Derivation of Strong Interactions from a Gauge Invariance}}, Nucl. Phys.{\bf{26}}, 222-229, (1962). \newline 
22. S. Gasiorowicz, op.\ cit., p.287. \newline
23.\ A. R. Edmonds, {\it{Angular Momentum in Quantum Mechanics}}, Princeton University Press, Princeton 1960, pp.64. \newline
24. M. E. Rose, {\it{Elementary Theory of Angular Momentum}}, (Dover Publications, New York 1995, John Wiley and Sons 1957), p.52. \newline
25. A. R. Edmonds, op. cit., p.65. \newline
26. M. E. Rose, op. cit., p.55. \newline
27. K. Nakamura et.\ al., op. cit., pp.72. \newline
28.\ S. R. Wadia,   $N = \infty$ {\it{Phase Transition in a Class of Exactly Solvable Model Lattice Gauge Theories}}, Phys. Lett. {\bf{B93}}, 403-410 (1980).\newline 
29. N. W. Ashcroft and N. D. Mermin, {\it{Solid State Physics}}, (Holt, Rinehart and Winston New York 1976), p.160. \newline
30. L. I. Schiff, op.\ cit. p.205.  \newline
31. http://dcwww.fys.dtu.dk/ \newline $\sim$trinham/BaryonLieProgrammes \newline
32. M. E. Rose, op.\ cit., p.52. \newline
33. M. E. Rose, op.\ cit., p.71. \newline
34.\ J. F. Cornwell, {\it{Group Theory in Physics}} (Elsevier Academic Press, Amsterdam, London, California, 1984/2004), Vol. 2, pp.466. \newline
35. I. J. R. Aitchison and A. J. G. Hey, {\it{Gauge Theories in Particle Physics}}, 2$^{\rm{nd}}$ ed. (Adam Hilger, Bristol and Philadelphia 1989), p.31. \newline
36.\ E. Klempt and J. M. Ricard, op.\ cit., p.1112, cf. M.Ablikim et.al, {\it{Partial wave analysis of}} 
$J/\psi$  {\it{to}} $p \overline{p}\pi^0$, Phys.Rev {\bf{D.80}}.052004; arXiv:0905.1562v4[hep-ex] 7 sep 2009. \newline
37.\ A. N. Aleev et al., {\it{Observation and Study of a Narrow State in a  
$ \Sigma^{-}(1,385){\rm{K}}^{+}$ 
System}}, Z. Phys. {\bf{C25}}, 205-212 (1984).\newline
38.\ L. Y. Zhu et al., {\it{Cross section measurements of charged pion photoproduction in hydrogen and deuterium from 1.1 to 5.5 GeV}}, Phys. Rev. {\bf{C71}}, 044603 (2005). \newline
39.\ See the Babar Collaboration: B. Aubert et.\ al, {\it{Measurement of the}} 
$B^0 \rightarrow \overline{\Lambda}p\pi^{-} $  
{\it{Branching Fraction and Study of the Decay Dynamics}}, Phys. Rev. {\bf{D79}}, 112009, (2009) \newline
40. B. Aubert et.\ al, {\it{Measurement of the}} 
$B^0 \rightarrow \overline{\Lambda}p\pi^{-} $  
{\it{Branching Fraction and Study of the Decay Dynamics}}, arXiv:hep-ex/060802v1 7Aug 2006.\newline
41. M. E. Rose, op.\ cit., p.58. \newline
42.\ S. Eidelman et.\ al (ParticleDataGroup), {\it{Review of Particle Physics}}, Phys. Lett. {\bf{B592}} (2004). \newline
43.\ A. Bettini, {\it{Introduction to Elementary Particle Physics}}, (Cambridge University Press, UK 2008), p.210. \newline
44. I. Madsen, {\it{Matematik 3, Kursus i Lie-grupper}}, (Lecture Notes in Danish, University of Aarhus, Denmark 1977), p.IV.1.2. \newline
45. V. Guillemin and A. Pollack, {\it{Differential Topology}}, (Prentice-Hall, New Jersey, USA 1974), p.163.  \newline
46. S. J. Brodsky, H.-C. Pauli and S. S. Pincky, op.\ cit., p.312. \newline
47.\ J. F. Donoghue, E. Golowich and B. R. Holstein, {\it{Dynamics of the Standard Model}} (Cambridge University Press, Cambridge 1992/1996) p.16. \newline
48. P. A. M. Dirac, {\it{The Principles of Quantum Mechanics}}, 4$^{\rm{th}}$ ed. (Oxford University Press 1989), p.144. \newline
49. L. I. Schiff, op.\ cit., p.211. \newline
50. L. I. Schiff, op.\ cit., p.236. \newline
51. L. Fonda and G.C. Ghirardi, op.\ cit., pp.468. \newline
52. S. Gasiorowicz, op.\ cit, p.267. \newline
53.\ R. R. Roy and B. P. Nigam, {\it{Nuclear Physics. Theory and Experiment}}, (John Wiley and Sons, New York 1967), p.231. \newline
54. L. I. Schiff, op.\ cit., p.241. \newline
55.\ Hans Bruun Nielsen, Technical University of Denmark (private communication 1997). \newline
56.\ T. Amtrup, {\it{Two integral presumptions}}, LMFK-bladet no. 4, April 1998. \newline 
57. A. Bettini, op.\ cit., p.204. \newline
58. I. Madsen, {\it{Matematik 3, Kursus i Lie-grupper}}, (Lecture Notes in Danish, University of Aarhus, Denmark 1977). \newline
59. V. Guillemin and A. Pollack, op.\ cit., p.177.\newline
60.\ A. O. Barut and R. Raczka, {\it{Theory of Group Representations and Applications}}, (World Scientific, Singapore 1986), p.86. \newline
61.\ M. R. Sepanski, {\it{Compact Lie Groups}}, (Springer, 2007), p.88. \newline
62. K. Nakamuru et.\ al., op.\ cit., p.204. \newline
63. I. Madsen, op.\ cit., p.III.2.10. \newline
64. I. Madsen, op.\ cit., p.III.1.7. \newline
65.\ Povl Holm, Rungsted Gymnasium, Denmark (private communication 1993). \newline

\end{document}